\documentclass[aps,prc,superscriptaddress,twoside,twocolumn,nofootinbib,10pt,showpacs,floatfix]{revtex4-1}
\usepackage{amsmath,amssymb}
\usepackage{graphicx,bm}

\usepackage{slashed}
\usepackage{epstopdf}
\usepackage{ulem} 
\usepackage[usenames]{color}
\usepackage{subfigure}
\usepackage{float}

\newcommand{\Slash}[1]{{\ooalign{\hfil/\hfil\crcr$#1$}}}

\newcommand{\rme}{ {\rm e} }
\newcommand{\rmd}{ {\rm d} }
\newcommand{\rmi}{ {\rm i} }

\newcommand{\lan}{ \langle }
\newcommand{\ran}{ \rangle }

\newcommand{\rp}{ {\rm p} }
\newcommand{\ra}{ {\rm a} }

\newcommand{\tq}{ \tilde{q} }

\newcommand{\vq}{ \vec{q} }
\newcommand{\vp}{ \vec{p} }
\newcommand{\vk}{ \vec{k} }

\newcommand{\calS}{ {\cal S} }

\newcommand \beq{\begin{eqnarray}}
\newcommand \eeq{\end{eqnarray}}

\usepackage[colorlinks=true,linktocpage=true,linkcolor=blue,citecolor=blue]{hyperref}

\allowdisplaybreaks

\begin{document}

\title{ Gluon propagator in two-color dense QCD: \\
Massive Yang-Mills approach at one-loop}

\author{Daiki Suenaga}
\email{suenaga@mail.ccnu.edu.cn}
\affiliation{Key Laboratory of Quark and Lepton Physics (MOE) and Institute of Particle Physics, Central China Normal University, Wuhan 430079, China}

\author{Toru Kojo}
\email{torujj@mail.ccnu.edu.cn}
\affiliation{Key Laboratory of Quark and Lepton Physics (MOE) and Institute of Particle Physics, Central China Normal University, Wuhan 430079, China}

\date{\today}

\newcommand\sect[1]{\emph{#1}---}
\begin{abstract}
We study the Landau gauge gluon propagators in dense two-color QCD at quark chemical potential, $\mu_q$, in the range from 0.5 to 1.0 GeV not reachable by the  perturbative method at weak coupling. In order to take into account the non-perturbative effects, at tree level we use a massive Yang-Mills model for the Yang-Mills theory (or the Curci-Ferrari model) which has successfully described the lattice results of the gluon and ghost propagators in the Landau gauge. We couple quarks to this theory and compute the one-loop polarization effects in medium. The presence of the gluon mass significantly tempers the medium effects and uncertainties associated with the strong coupling constant $\alpha_s$. The diquark condensate in two-color QCD is color-singlet, for which neither electric nor magnetic screening masses should appear at the scale less than the diquark gap. The presence of the gap helps to explain the lattice results which are not very sensitive to the quark density. Meanwhile we also found the limitation of the one-loop estimate as well as the lack of some physics in perturbative medium corrections.

\end{abstract}

\maketitle

\section{Introduction}

A highly compressed matter of quantum chromodynamics (QCD) is expected to transform from a hadronic to a quark matter when baryons overlap; then quarks (and gluons) start to directly contribute to equations of state as well as transport properties of the matter \cite{Baym:1976yu}. Considering the size of hadrons of $\sim 0.5$-$1\, {\rm fm}$ the transition should occur around the baryon density $ n_B \sim 5$-$10\, n_0$ ($n_0 \simeq 0.16 \, {\rm fm}^{-3}$: nuclear saturation density) or quark chemical potential $\mu_q = 0.5$-$0.8\, {\rm GeV}$ \cite{Baym:2017whm,Baym:2019iky}. Such dense matter may be realized at the cores of the two-solar mass neutron stars discovered in binary systems \cite{Demorest:2010bx,Fonseca:2016tux,Antoniadis:2013pzd} including the most recent one with the mass $2.14\pm0.10$ solar mass at $68.3\%$ confirmation level \cite{Cromartie:2019kug}.

The direct QCD calculations for quark matter have been based on the perturbation theory and carried out to 3-loop order \cite{Freedman:1976ub,Fraga:2001id,Kurkela:2009gj}. But these calculations at $\mu_q \lesssim 1\,{\rm GeV}$ or $n_B \lesssim 50 \, n_0$ show that the perturbative series do not converge well \cite{Fraga:2001id}, or the renormalization scale dependence is large \cite{Kurkela:2009gj}. These QCD calculations, together with the estimate of the onset density of quark matter, suggest that matter at $\mu_q = 0.5$-$1\, {\rm GeV}$ is strongly correlated quark matter \cite{Baym:2017whm}. In order to explore this region one needs to develop a framework based on quarks and gluons but must retain strong coupling effects.

\begin{figure}[b]
\vspace{-0.2cm}
\centering
\includegraphics[scale=0.25]{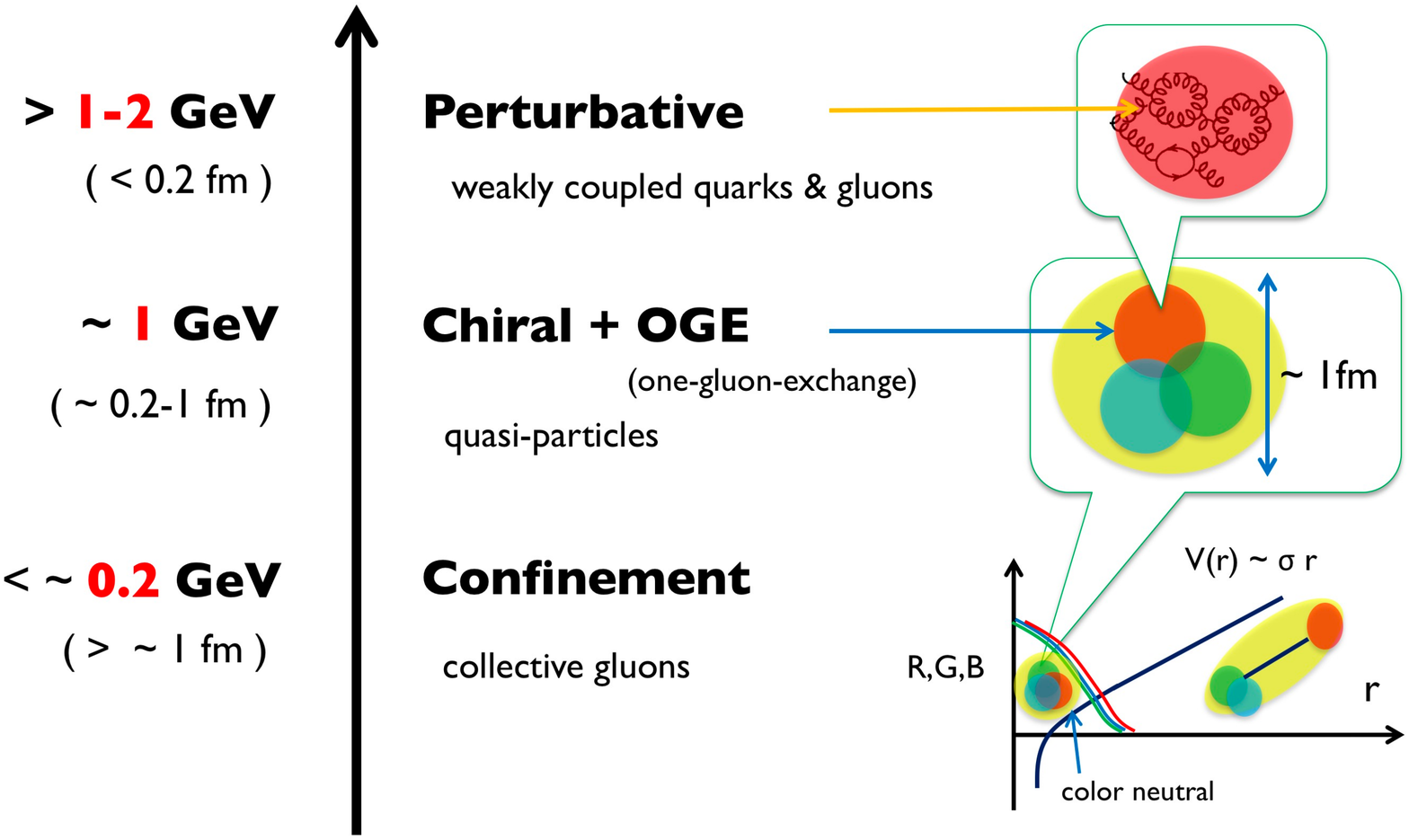}
\vspace{-0.3cm}
\caption{
A schematic description of a single hadron. The vertical axis represents the energy (distance) scale.
  }
  \vspace{-0.cm}
   \label{fig:3window_hadrons}
\end{figure}

Concerning the strong coupling effects at finite density, the theoretical description of the confinement-deconfinement phenomenon remains a difficult problem, see \cite{Greensite}  for various theoretical scenarios. But for a matter dense enough for the color-singlet state to appear locally and homogeneously, the detailed account of confining forces might not be so important for most of physical quantities, except colored excitations on top of the color-white background. This is the case for spatially one-dimensional QCD in which the color-flux remains confining from low to high densities nevertheless the equation of state is dominated by free quark gas contributions \cite{Schon:2000he,Bringoltz:2008iu,Bringoltz:2009ym,Kojo:2011fh}. Inspired by this result, we conjecture that, in the domain where the color-singlet condition is satisfied, the quasi-particle picture for quarks and gluons can be applied at distance of $\lesssim 1\, {\rm fm}$ or momentum transfer of $0.2$-$1\, {\rm GeV}$, as in the constituent quark models where quarks with effective chiral masses of $M_q=300$-$500$ MeV explain the dynamics inside of hadrons \cite{DeRujula:1975qlm}, see a schematic picture in Fig.\ref{fig:3window_hadrons}. For the quasi-particle descriptions to be useful, the strong coupling effects should be largely absorbed into the effective mass, coupling, and so on, after which the residual interactions should be under control \cite{Manohar:1983md,Weinberg:2010bq}.

This paper is our first step to the quasi-particle description for strongly correlated quark matter and we take up 2-color QCD (QC$_2$D) as a testing ground. In this theory the lattice QCD simulation is possible without suffering from the sign problem and one can confront his calculations with the lattice data for the phase diagram, equations of state, diquark condensates, Polyakov loops, and so on \cite{Hands:2010gd,Cotter:2012mb,Braguta:2016cpw}. Also the Landau gauge gluon propagators and vertices have been measured \cite{Hajizadeh:2017ewa,Boz:2018crd}. For model studies of QC$_2$D, see Ref.\cite{Strodthoff:2011tz} and work in the context of quarkyonic matter \cite{McLerran:2007qj}, see Ref.\cite{Brauner:2009gu}.

In this work we study the in-medium modification of the Landau gauge gluon propagator, including quark loop effects to one-loop. We combine the in-medium effects with non-perturbative vacuum gluon propagators. For the latter, the Landau gauge studies in lattice QCD \cite{Cucchieri:2011ig,Maas:2014xma} and functional approaches \cite{vonSmekal:1997ern,Alkofer:2006jf,Fischer:2008uz,Cyrol:2016tym} for pure Yang Mills (YM) theory have reported the generation of effective mass of $m_g \sim 0.4$-$0.7\, {\rm GeV}$ at soft Euclidean momenta (see Refs.\cite{Parisi:1980jy,Cornwall:1981zr,Mandula:1987rh} for early studies). Based on this finding seminal works assumed the massive Landau gauge YM, or the Curci-Ferrari (CF) model \cite{Curci:1976bt} as an effective theory and performed the 1-loop calculations for gluon and ghost propagators, finding the remarkable agreement with the lattice results in vacuum \cite{Reinosa:2017qtf,Tissier:2011ey} and reasonable agreement at finite temperature \cite{Reinosa:2013twa,Reinosa:2016iml}. Encouraged by these findings, we use the gluon and ghost propagators in the CF model  
as our tree level propagators, and add the polarization effects due to quarks in medium. For vacuum gluon and ghost propagators with dynamical quarks, see Ref.\cite{Pelaez:2014mxa} for the CF model, 
Ref.\cite{Bowman:2007du} for the lattice results, and Ref.\cite{Cyrol:2017ewj} for the results of functional calculations. Also, the CF model was applied at nonzero chemical potential including (heavy) quarks to investigate the QCD phase diagram~\cite{Reinosa:2015oua,Maelger:2017amh,Maelger:2018vow,Maelger:2019cbk}.

The analyses of in-medium gluon propagators, however, can in principle be more non-linear and complex, as the quark loop effects may strongly depend on the phase structure \cite{Rischke:2000qz,Rischke:2000ra,Huang:2004bg,Huang:2004am,Fukushima:2005cm,Kojo:2014vja}. For example the quarks entering the loop can be either gapped or gapless depending on the pairing near the Fermi surface, and add totally different contributions to the gluon polarization functions. Following the previous one-loop study \cite{Kojo:2014vja}, we classify three distinct possibilities of phases and the corresponding screening mass effects: (i) {\it normal} phase, in which quarks are gapless. Here the gluons acquire the electric mass from gapless particle-hole excitations, but no magnetic mass, due to the exact cancellation between the paramagnetic contribution (due to the particle-hole) and diamagnetic contribution (due to the particle-antiparticle); (ii) {\it Higgs} phase, in which quark-pairs form a colored diquark condensate and quarks are gapped, while the phase fluctuations of the condensate are colored and hence couple to the longitudinal mode of gluons, yielding both electric and magnetic (Meissner) masses; (iii) {\it singlet} (gapped) phase, in which the diquarks form a color-singlet condensate and quarks are gapped, while the color-singlet phase fluctuations of the condensate do not couple to gluons. In this case the gapped quarks and the absence of Meissner effects together protect gluons from acquiring electric and magnetic masses. In this paper we investigate the normal and singlet phases of QC$_2$D, using the CF model.

The singlet phase corresponds to the BCS phase in QC$_2$D where the most favorable pairing is anti-symmetric with respect to color, flavor, and spin, while the spatial wavefunction is S-wave. The lattice calculations found that the critical temperature is $T_c \simeq 80$-$120\, {\rm MeV}$ so we estimate diquark gaps $\Delta$ to be $140$-$210\, {\rm MeV}$ by assuming the BCS formula $T_c \simeq 0.57 \Delta$. Since this matter is an insulator, the gluons are unscreened at scale lower than $\sim \Delta$. This observation is consistent with the recent lattice results for QC$_2$D at $\mu_q =0.5$-$1$ GeV \cite{Boz:2018crd}, where the electric and magnetic gluon propagators, ghost propagators, and gluon-ghost vertices are not as sensitive to the variation of $\mu_q$ as predicted by the normal phase scenario. 

In our analyses for QC$_2$D we do not manifestly calculate the diquark gaps $\Delta$, but just treat them as given in the range of 0-200 MeV. Then we use quark propagators with $\Delta$ to compute the polarization effects. As we will see $\Delta$ improves the consistency with the lattice results in the electric sector. Quantitatively, the overall size of the quark loop strongly depends on the strong coupling $\alpha_s$ and our choice of the renormalization scale for it; a proper renormalization scale should be used to minimize truncation errors in practical diagrammatic calculations. 
In the infrared its value can be as large as $\sim 3$ (see Ref.\cite{Deur:2016tte} for the recent summary about $\alpha_s$ extracted in various approaches). 
Although the CF model is provided as an effective theory of QCD at the infrared regime, the value of coupling within the CF model still includes an uncertainty. For this reason, we vary it considerably, from 0 to 3, to cover wide range of possibilities. 
It turns out, however, that the presence of the gluon mass in the vacuum propagator largely tempers the impact of varying $\alpha_s$. Similar observation was made for the hot QCD equations of state in Ref.\cite{Fukushima:2013xsa}, where the authors applied the Gribov-Zwanziger gluon propagators. We also expect that the insensitivity to $\alpha_s$ should significantly stabilize our analyses of various quantities at finite density. This is one of the main conclusions in this work.

Our calculations of the polarization functions maintain the conservation law or symmetry by handling the regularization artifacts which need special care. 
The regularization artifacts would appear when (i) there are non-perturbative changes in fermion bases, from the vacuum ones to the medium ones; and (ii) the gaps are made constant from the IR to the UV regions. These features are typical for in-medium computations in practice. Such UV artifacts can couple to the medium effects in the IR, yielding unphysical screening masses which are UV finite. In the previous study \cite{Kojo:2014vja}, the authors cancel the symmetry violating UV artifacts with symmetry violating counter terms, by demanding the final expression to recover the conservation laws. In this paper we offer a simpler method in which the conservation law is kept at every step of computations so that one needs only the standard set of counter terms.

This paper is structured as follows. In Sec.\ref{sec:Model} we summarize our models for gluons and the possible pairing patterns. In Sec.\ref{sec:SelfEnergy_general} we discuss general remarks on the polarization functions, and in Sec.\ref{sec:regularization} explain how to preserve the conservation laws during computations. In Sec.\ref{sec:OneLoop} we present our one-loop results and compare them with the lattice data. In Sec.\ref{sec:pheno_fit} is devoted to discussions about the non-perturbative considerations beyond one-loop. Sec.\ref{sec:summary} is devoted to the summary.

We use the following notations: $\int_q \equiv \int \rmd^4 q/(2\pi)^4$, $\int_{\vq} \equiv \int \rmd^3 \vq/(2\pi)^3$. The matrices $\sigma^a (a=1,2,3)$ and $\tau^i (i=1,2,3)$  are the Pauli matrices with respect to the color and flavor spaces, respectively. We freely raise or lower the color and flavor indices when the notations become simpler. As for the space-time metric, we will work on the Euclidean space but we leave the upper and lower indices as in the Minkowski space. In this way we can transfer the expression developed in the Minkowski space to the Euclidean with minimal efforts. The relations $k_\mu = g_{\mu \nu} k^\nu$, $a\cdot b = g_{\mu \nu} a^\mu b^\nu$, 
$\{\gamma_\mu, \gamma_\nu \} = 2 g_{\mu \nu}$, are common for these two spaces. In the components  $g_{\mu \nu}^E = - \delta_{\mu \nu} $, $g_{\mu \nu}^M = (1, -1,-1,-1)_{{\rm diag}}$, $a^E_4 = -\rmi a^M_0 (a_E^4 = \rmi a_M^0)$, $a^E_j = a^M_j$ for four-vectors, and $\gamma^E_\mu = - ( \gamma^E_\mu )^\dag$. The only difference we should care is the overall factor of propagators ${\cal G}$ as $ -\rmi {\cal G}_M (k) = {\cal G}_E (k) $, and the $\rmi \epsilon$ term in the Minkowski expression. The others need not be modified. When we emphasize the positivity of the scalar product of momenta, we occasionally use the capital letters, e.g., $K^2 = - k^2 (\ge 0)$, and also the notation $K=\sqrt{-k^2}$. The convention for the self-energy $\Pi$ is $D^{-1} = D_{\rm tree}^{-1} + \Pi$ where $D$ and $D_{\rm tree}$ are the dressed and tree level propagators.

\section{Model} \label{sec:Model}

\subsection{A model for gluons} \label{sec:Model_gluon}

For quasi-particle descriptions for gluons we use a model introduced by Curci-Ferrari \cite{Curci:1976bt}, which is the massive YM Lagrangian with the Landau gauge fixing condition \cite{Tissier:2011ey,Reinosa:2017qtf}. The theory is renormalizable with finite set of counter terms \cite{Curci:1976bt}, as in the pure YM theory. On the other hand this model does not preserve the perturbative unitarity \cite{Curci:1976kh,deBoer:1995dh}.
Whether the unitarity is recovered or not in a nonperturbative regime, see discussions in Ref.\cite{Kondo:2012ri}.

We will regard that the gluon mass emerges from the dynamics in the Landau gauge. Hence, even though the introduction of the gluon mass already breaks the gauge invariance, we use the massive YM Lagrangian together with terms that enforce the Landau gauge fixing condition, $ \partial^\mu A_{\mu}^{ a} =0$. Now the Lagrangian for the CF model is
\begin{eqnarray}
{\cal L}_{\rm gauge}
&=& - \frac{1}{\, 4 \,} G^{\mu\nu}_a G_{\mu\nu}^{ a} + \frac{\, m_{g}^2 \,}{2}  A_a^{\mu } A_\mu^a \nonumber\\
&& - \frac{1}{\, 2 \alpha \,} \left( \partial^\mu A_{\mu}^{ a} \right)^2
+ \bar{c}^a \rmi\partial^\mu D_\mu c^a \,,
\label{MassiveYM}
\end{eqnarray}
where $A_\mu^a$ and $c^a$ are the gluon and ghost fields, respectively, with the color indices $a=1,2,3$ (for QC$_2$D), and $m_g$ is the gluon mass. For the moment we keep the gauge parameter $\alpha$ in the Lagrangian but in the end we will take the limit $\alpha \rightarrow 0$.
The covariant field strength $G_{\mu\nu}^a$ is defined by
\begin{eqnarray}
G_{\mu\nu}^a = \partial_\mu A_\nu^a-\partial_\nu A_\mu^a + g f^{abc}A_\mu^b A_\nu^c\ ,
\end{eqnarray}
where $g$ is the gauge coupling constant and $f^{abc}$ is the structure constant. The covariant derivative for the ghost field $c^a$ is
\begin{eqnarray}
D_\mu c^a &=& \partial_\mu c^a + g f^{abc} A_\mu^b c^c \ .
\end{eqnarray}
The resultant tree level propagator, after putting $\alpha \rightarrow 0$, is (in Euclidean space)
\beq
\left[ D^{ab}_{ \mu\nu}(k) \right]_{ {\rm tree} }= D_{ {\rm tree} } (k) \delta^{ab} P_{\mu \nu}  \,, ~~ P_{\mu\nu} = g_{\mu\nu}- \frac{k_\mu k_\nu}{k^2}  \,, \nonumber\\
\eeq
which is transverse, $k^\mu P_{\mu \nu}=0$, and
\beq
D_{ {\rm tree} }(k) = \frac{-1}{\, k^2 - m_g^2 \,} \,.
\eeq
The radiative corrections in the CF model 
may contain the radiative corrections which are not transverse, but thanks to the Landau gauge condition its longitudinal component anyway can be dropped off from the dressed gluon propagators.

We regard this tree level Lagrangian as the consequence of non-perturbative calculations. Thus the suitable choice of the tree level mass $m_g$ can differ for theories with and without quarks. We will come back to this point after performing one-loop calculations with quarks.

\subsection{A model for quarks} \label{sec:Model_gluon}

In order to examine the pairing effects we use an effective Lagrangian in which diquark operators couple to the gap parameters $\Delta$. Such gaps are produced by diquark condensates for which one can consider several quantum numbers. For the Dirac mass associated with the chiral symmetry breaking (ChSB) we use the effective quark mass $M_q$ rather than the current quark mass $m_q$. In this paper we will not solve the gap equations to derive $\Delta$ and $M_q$, but simply choose some characteristic values to examine the impact of medium effects. 

We usually guess the most favorable diquark pairing by applying the one-gluon exchange picture. But its validity is uncertain at strong coupling. So we also present another qualitative description here. The condensate should be color-antisymmetric, as it reduces its color-charge and the associated color-electric flux (in QC$_2$D, such a diquark condensate is color-singlet). Then, the flavors, spins, and spatial wavefunctions should form a symmetric wavefunction as a total. For the spatial wavefunction the S-wave pairing should be most preferable as one can fully utilize the entire Fermi surface for quark pairing. For the flavor wavefunction, we assume it to be anti-symmetric as the system can reduce the flavor charges; accumulation of charges usually produce fields and cost more energy. Taking all these considerations the condensate should be spin-singlet, leading to the form
\begin{eqnarray}
\langle \psi^T C\gamma_5 \sigma^2\tau^2\psi\rangle ~ \sim ~ \Delta \mu_q^2  \,,
\label{GapSU2}
\end{eqnarray}
where the matrices $\sigma^2$ and $\tau^2$ combines the color and flavor indices of quarks in antisymmetric way. $C$ is the charge-conjugation matrix defined by $C=-\gamma_2\gamma_4$. The factor $\mu_q^2$ comes from the phase space near the Fermi surface, $\sim 4\pi \mu_q^2$, at large density.

The diquark condensate in Eq.~(\ref{GapSU2}) is not invariant with respect to $U(1)_B$-transformations, $\psi \rightarrow \rme^{ \rmi \theta} \psi$, but invariant with repsect to the $SU(2)_c$-transformations, $\psi \rightarrow \rme^{ \rmi \theta_a \sigma_a/2} \psi$, as the condensate does not carry color charges. Thus unlike the color-superconductivity in 3-color QCD, the phase fluctuations of diquark condensates do not participate in the longitudinal modes of gluons, and hence no Messner mass is generated. The phase fluctuations simply appear as gapless Nambu-Goldstone modes associated with the $U(1)_B$ symmetry breaking.

The effective Lagrangian for quarks takes the form
\begin{eqnarray}
{\cal L}_\psi = \bar{\psi}( \rmi \Slash{D} + \rmi \mu_q\gamma_4- M_q)\psi - \psi^T {\bf \Delta} \psi\ ,  \label{QLagrangian}
\end{eqnarray}
where $M_q$, $\mu_q$, and ${\bf \Delta} \equiv  \sigma^2\tau^2\gamma_5 \Delta$ are the effective quark mass, quark chemical potential, and a matrix for the diquark gap, respectively. The covariant derivative is
\begin{eqnarray}
D_\mu \psi &=& \partial_\mu \psi + \rmi g A_\mu^a \frac{\sigma^a}{2} \psi \,.
\end{eqnarray}
The standard technique to handle the mean field di-fermion condensate is the Nambu-Gor'kov formalism. Useful summary can be found in Ref.\cite{Rischke:2000qz}. Introducing a two-component spinors
\begin{eqnarray}
\Psi \equiv \frac{1}{\sqrt{2}}\left(
\begin{array}{c}
\psi \\
\psi_c \\
\end{array}
\right)\ , \ \bar{\Psi} \equiv \frac{1}{\sqrt{2}}\left(\bar{\psi}\,,  \bar{\psi}_c \right) \,,
\end{eqnarray}
and using a relation $\psi^T =-\bar{\psi}_c C$ ($\bar{\psi}^T=-C\psi_c$), the Lagrangian~(\ref{QLagrangian}) is rewritten into
\begin{eqnarray}
{\cal L}_q = \bar{\Psi}{\cal K} \Psi - g\bar{\Psi}\Slash{\cal A}\Psi\ ,
\end{eqnarray}
where we have defined a matrix for the quark bilinear terms, 
\begin{eqnarray}
{\cal K} = \left(
\begin{array}{cc}
\rmi \Slash{\partial} + \rmi \mu_q \gamma_4 - M_q & \bar{\bf \Delta} \\
{\bf \Delta} & \rmi \Slash{\partial} - \rmi \mu_q\gamma_4 - M_q \\
\end{array}
\right)\ , \label{KineticQ}
\end{eqnarray}
($\bar{\bf \Delta} = \gamma_0{\bf \Delta}^\dagger\gamma_0$) and the bare vertex matrix
\begin{eqnarray}
\Slash{\cal A} &=& \gamma_\mu^a {A}_\mu^a \,,~~~~\gamma_\mu^a = \gamma_\mu R^a \,,
\end{eqnarray}
with
\begin{eqnarray}
R^a \equiv \left(
\begin{array}{cc}
\sigma^a/2 & 0 \\
0 & -(\sigma^a)^T/2\\
\end{array}
\right) \ . 
\label{BareVertex}
\end{eqnarray}
Next, we construct a tree level propagator from the quark bilinear term. According to Eq.~(\ref{KineticQ}), the inverse of the propagator reads in the momentum space 
\begin{eqnarray}
{\cal S}^{-1}(\tq) = \left(
\begin{array}{cc}
\Slash{q} + \rmi \mu_q \gamma_4 - M_q  & \bar{\bf \Delta} \\
{\bf \Delta} & \Slash{q} - \rmi \mu_q \gamma_4 - M_q \\
\end{array}
\right) \,. \nonumber\\
 \label{NSInv}
\end{eqnarray}
Below we assume the diquark gap function ${\bf \Delta}$ to be constant. More realistically it should be vanishing for quarks away from the Fermi surface. In order to find the expression for $\calS$, it is convenient to decompose the matrix into the particle and antiparticle components. We introduce the particle (p) and antiparticle (a) projection operators
\begin{eqnarray}
\Lambda_{\rp,\ra} = \gamma_0 \frac{E_{q}\gamma_0 \pm \left( M_q+\vec{\gamma}\cdot \vec{ q} \right)}{2E_{ q}} \,,
\end{eqnarray}
where $E_{ q } = \sqrt{ \vec{q}^2 + M_q^2}$. We also express $\Delta$ as 
\begin{eqnarray}
\Delta = \Delta \Lambda_\rp + \Delta \Lambda_\ra \,.
\end{eqnarray} 
Solving an equation $\calS \calS^{-1} ={\bf 1}$, we find the quark propagator of the form
\begin{eqnarray}
\calS = \left(
\begin{array}{cc}
S_{11}^D & \tau^2 \sigma^2 S_{12}^D \\
\tau^2 \sigma^2 S_{21}^D & S_{22}^D \\
\end{array}
\right)\ ,
\end{eqnarray}
with ($\Lambda_\rp^C =\Lambda_\ra$ and $\Lambda_\ra^C = \Lambda_\rp$)
\begin{eqnarray}
\calS_{11}^D 
&=& 
 \left( \frac{|u_\rp|^2}{\rmi q_4 - \epsilon_\rp} + \frac{|v_\rp|^2}{ \rmi q_4 + \epsilon_\rp}\right) \Lambda_\rp \gamma_0 
\nonumber\\
&&
\ \ \ + \left(\frac{|v_\ra|^2}{ \rmi q_4 - \epsilon_\ra} + \frac{|u_\ra|^2}{\rmi q_4 + \epsilon_\ra}\right) \Lambda_\ra \gamma_0 
\nonumber\\
\calS_{12}^D 
&=& 
- \left( \frac{u_\rp^* v_\rp^*}{ \rmi q_4 - \epsilon_\rp} - \frac{u_\rp^* v_\rp^*}{ \rmi q_4 + \epsilon_\rp}\right) \Lambda_\rp \gamma_5 
\nonumber\\
&&
\ \ \ - \left( \frac{u_\ra^* v_\ra^*}{ \rmi q_4 - \epsilon_\ra} - \frac{u_\ra^* v_\ra^* }{ \rmi q_4 + \epsilon_\ra} \right) \Lambda_\ra \gamma_5 
\nonumber\\
\calS_{21}^D 
&=&
  \left( \frac{u_\rp v_\rp}{ \rmi q_4 - \epsilon_\rp} - \frac{u_\rp v_\rp}{\rmi q_4 + \epsilon_\rp}\right) \Lambda_\rp^C \gamma_5 
 \nonumber\\
&&
\ \ \ + \left( \frac{u_\ra v_\ra}{\rmi q_4 - \epsilon_\ra} - \frac{u_\ra v_\ra}{\rmi q_4 + \epsilon_\ra}\right) \Lambda_\ra^C \gamma_5
 \nonumber\\
\calS_{22}^D 
&=& \left( \frac{|v_\rp|^2}{ \rmi q_4 - \epsilon_\rp} + \frac{|u_\rp|^2}{ \rmi q_4 + \epsilon_\rp}\right) \Lambda_\rp^C \gamma_0
 \nonumber\\
&&\ \ \  + \left( \frac{|u_\ra|^2}{ \rmi q_4 - \epsilon_\ra} + \frac{|v_\ra|^2}{ \rmi q_4 + \epsilon_\ra} \right) \Lambda_\ra^C \gamma_0  \ , 
\label{QPropagatos}
\end{eqnarray}
where $\epsilon_p$, $\epsilon_a$ are quasi-particle dispersions, 
\begin{eqnarray}
\epsilon_\rp 
&=& \sqrt{(E_q-\mu_q)^2+|\Delta|^2} 
\nonumber\\
\epsilon_\ra 
&=& \sqrt{(E_q+\mu_q)^2+|\Delta|^2} \,,
\end{eqnarray}
and $u_\rp$, $v_\rp$, $u_\ra$, $v_\ra$ are factors satisfying the following relations:
\begin{eqnarray}
|u_\rp|^2 = \frac{1}{2} \left(1+\frac{E_q - \mu_q}{\epsilon_\rp } \right)\ &,&\ \  |u_\ra|^2 = \frac{1}{2} \left(1+\frac{E_q+\mu_q}{\epsilon_\ra}\right) \,, \nonumber\\
|v_\rp|^2 = \frac{1}{2} \left(1-\frac{E_q - \mu_q }{\epsilon_\rp }\right)  \ &,&\ \  |v_\ra|^2 = \frac{1}{2} \left(1-\frac{E_q+\mu_q}{\epsilon_\ra}\right) \,, \nonumber\\
\end{eqnarray}
and
\begin{eqnarray}
&&|u_\rp |^2 + |v_\rp|^2 = |u_\ra |^2 + |v_\ra |^2=1 \,, \nonumber\\
&& u_\rp v_\rp = \frac{\Delta}{2\epsilon_\rp} \ ,\ \ u_\ra v_\ra  =\frac{\Delta }{2\epsilon_\ra}\ .
\end{eqnarray}
%

\section{Self-energy: general remarks}
\label{sec:SelfEnergy_general}

In this section we give general remarks on the structure of the gluon self-energy and a new renormalization condition which is associated with the gluon mass term in the tree level Lagrangian. We first review the treatment for the gluon self-energy 
of the CF model for the pure YM theory  
($\Pi^{ {\rm YM} }$), and then include quarks in vacuum ($\Pi^{ {\rm vac} }$). Finally we discuss general remarks on the gluon self-energy in-medium ($\Pi^{ {\rm} }$).

\subsection{Vacuum cases}
\label{sec:vac}

Unlike the massless YM theory, the modified Ward-Takahashi (WT) identity for the CF model 
leads to the gluon self-energy tensor which includes the terms proportional to $m_g^2 g_{\mu \nu}$. 
They contribute to the transverse as well as the longitudinal components,
\begin{eqnarray}
\Pi^{ {\rm YM} }_{\mu\nu}(k) = \Pi^{ {\rm YM} } (k) P_{\mu\nu} + \Pi_L^{ {\rm YM} } (k) \frac{\, k_{\mu} k_{\nu} \,}{k^2} \,.
\end{eqnarray}
But we use the massive YM theory 
together with the Landau gauge condition; as we have already mentioned the longitudinal component decouples from the gluon propagator. Hereafter we discuss only the transverse part.

Now we discuss how to handle the UV divergences in the CF model.
For the vacuum computation we use the dimensional regularization as it satisfies the 
WT identity. Then it is guaranteed that the divergences specific to the CF model 
appear as the coefficient of $m_g^2$ terms and are at most logarithmic. Such logarithmic divergence can be cancelled by a new mass counter term which originates from the gluon mass term. Now the renormalized self-energy includes the bare (regularized) function and counter terms,
\beq
\Pi_{ {\rm YM} }  (k) = \Pi_{ {\rm YM} }^{ {\rm bare} } (k) - k^2 \delta_{Z_g}^{ {\rm YM} } + \delta^{ {\rm YM} } m_g^{2 } \,.
\eeq
Here we have two counter terms and hence we must set up two renormalization conditions. Following Ref.\cite{Tissier:2011ey}, we choose our renormalization points to be 
\begin{eqnarray}
\Pi_{ {\rm YM} } (\mu _{{\rm R}}  ) =  \Pi_{ {\rm YM} } (0 ) = 0 \,,
\label{SU2GluonRC}
\end{eqnarray}
where $\mu_{ {\rm R} }$ is some renormalization points.

Next we include quarks. They do not change the structure of the gluon self-energy. So we have only to make replacements,
\beq
\left( \Pi,\delta_{Z_g}, \delta m_g^2, \mu_R \right)^{ {\rm YM} }  \rightarrow \left( \Pi,\delta_{Z_g}, \delta m_g^2, \mu_R \right)^{ {\rm vac} } \,.
\eeq
%

\subsection{In-meidum self-energy}
\label{sec:MediumSelfEnergy}

In medium, the presence of matter breaks the Lorentz symmetry and one must deal with electric and magnetic components differently. Then the projector $P_{\mu \nu}$ splits into
\beq
P_{\mu \nu} = P^E_{\mu\nu} + P^M_{\mu\nu}  \,,
\eeq
where the projector for magnetic components is three-dimensionally transverse,
\beq
&& P_{44}^M = P_{i0}^M=P_{0i}^M = 0\ ,
~~ P_{ij}^M = - \delta_{ij} + \frac{k_ik_j}{|\vec{k}|^2} \ , 
\label{PM}
\eeq
so that the electric tensor is
\beq
P_{\mu \nu}^E = P_{\mu \nu} -P^M_{\mu\nu}\ . 
\label{PE}
\eeq
Using these projectors the polarization tensor in medium can be written as
\beq
\Pi_{\mu\nu}(k) = \Pi_{E}(k)P^E_{\mu\nu} + \Pi_{M}(k)P^M_{\mu\nu} + \Pi^L_{\mu \nu}  (k) \,.
\eeq
Therefore the Landau gauge gluon propagators must be of the form ($K^2 = -k^2 \ge 0)$
\begin{eqnarray}
D_{\mu\nu}(k)  = \sum_{i=E,M} \frac{1}{K^2+m_g^2+\Pi_i (k)}P^i_{\mu\nu}  \,,
\label{PropagatorSU2}
\end{eqnarray}
where as we have already mentioned $\Pi^L_{\mu \nu} (k)$ could be dropped off because it does not couple to the tree Landau gauge propagator.

The counter terms set up in vacuum will be also used in medium computations for $\Pi_{E,M}$. We make the decomposition for the in-medium polarization function (for the moment we suppress the subscript $E$ and $M$), 
\beq
\Pi = \Pi_{ {\rm vac} } + \delta \Pi\,,~~~~ \delta \Pi =  \Pi^{ {\rm bare} } - \Pi_{ {\rm vac} }^{ {\rm bare} }  \,,
\label{PiRelation}
\eeq
where the counter terms are already included into $ \Pi_{ {\rm vac} }$ while $\delta \Pi$ includes the difference between the bare self-energies.  Below we focus on the term $ \delta \Pi $.

The term $ \delta \Pi $ would look insensitive to the UV contributions. The medium part is most typically computed by picking up the residues with an implicit assumption of the three-dimensional cutoff regularization, $|\vp| \le \Lambda_{ {\rm UV} }$ (otherwise poles may exist outside of the contour in the complex $p_0$-plane). But computations of $ \delta \Pi $ with such regularization would suffer from UV artifacts that violate the 
WT identity. It impacts on the qualitative behaviors of gluon self-energy; without removing this artifact the gluons would acquire spurious magnetic masses.

Without satisfying the WT identity, each of $\Pi^{ {\rm bare} }$ in the medium and in vacuum has the quadratic divergence and the single subtraction, $\Pi^{ {\rm bare} } - \Pi_{ {\rm vac} }^{ {\rm bare} } $, leaves terms that couple to the differences between quark bases (in the UV domain). Such terms are absent in the regularization consistent with the WT identity; in that case the leading divergences before the subtraction are at most logarithmic and hence the difference between the bases appear as coefficients of $\Lambda^{-2}_{ {\rm UV} }$. Since we are trying to go beyond the perturbative framework by including the modification of quark bases, this problem deserves special remarks. In the next section we will introduce a practical scheme which is free from the artifacts.

\section{In-medium regularization}
\label{sec:regularization}


\begin{figure}[bp]
\centering
\vspace{-1.0cm}
\includegraphics[scale=0.2]{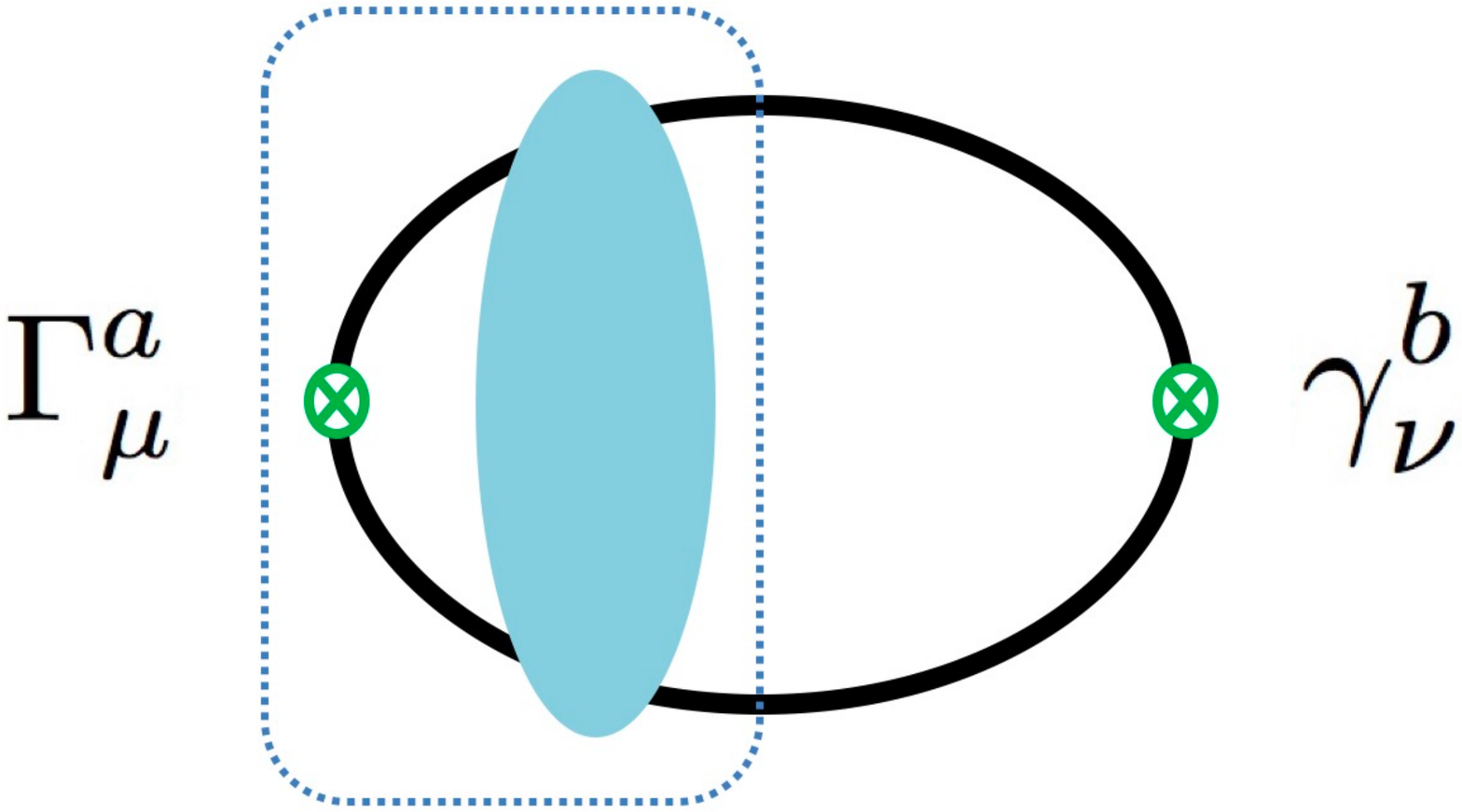}
 \vspace{-1.0cm}
 \caption{(color online) The diagrammatical picture of the current-current correlator $\left(\Pi_q^{\rm bare}\right)_{\mu\nu}^{ab}$ in Eq.(\ref{eq:pol}). 
 The Abelian-type vertex circled by a blue curve ($\Gamma_\mu^a$) is shown in Fig.~\ref{fig:VertexA}, and the solid lines are full quark propagators.
  }
\label{fig:current-current}
\end{figure}

\begin{figure}[bp]
\centering
\includegraphics[scale=0.22]{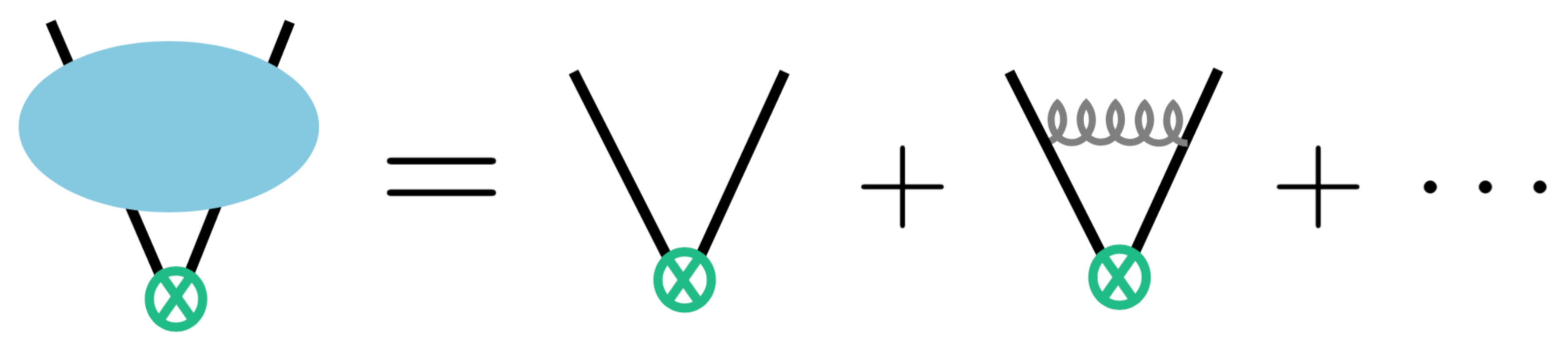}
 \caption{(color online) The Abelian-type vertex $\Gamma_\mu^a$.
  }
\label{fig:VertexA}
\end{figure}

\begin{figure}[htbp]
\centering
\includegraphics[scale=0.2]{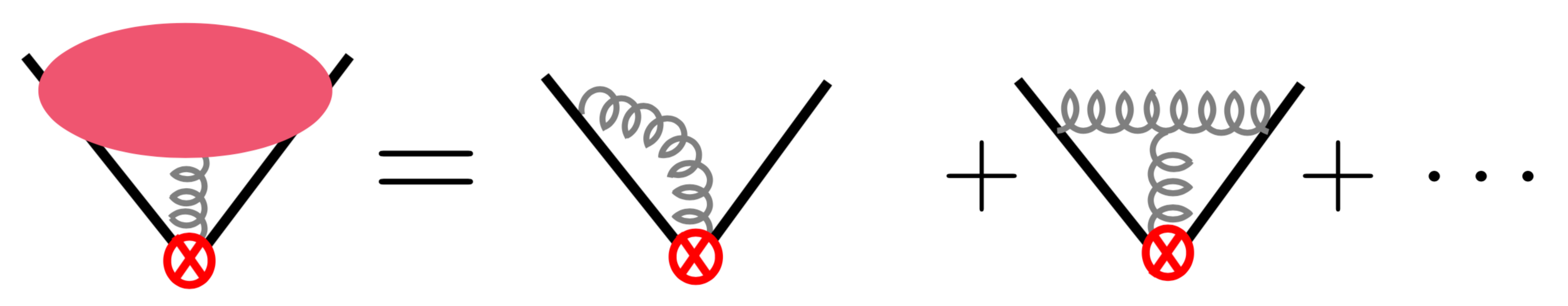}
 \caption{(color online) The non-Abelian-type vertex $L_{\cal NA}^a$ only appearing in the WT identity.} 
\label{fig:VertexNA}
\end{figure}

In this work we consider the medium effects which come from a current-current correlator (Fig.\ref{fig:current-current}),
\beq
\left(\Pi^{\rm bare}_q (k) \right)_{\mu\nu}^{ab}= - g^2 \int_x \rme^{-\rmi k x} \langle j^a_\mu (x)  j^b_\nu (0) \rangle \label{JJCorrelator}
\eeq
where $j_\mu^a = \bar{\psi} \gamma_\mu^a \psi$. Defining a general three point vertex function $\Gamma^a_\mu$ as (Fig.\ref{fig:VertexA})
\beq  
&\left\lan \,  j^a_\mu (x) \Psi(z_1) \bar{\Psi}(z_2) \,\right\ran \nonumber\\
& \equiv \int_{w,u} \calS(z_1-w)\, \Gamma^a_\mu(w-x, x-u) \, 
\calS (u-z_2) \,,
\label{JPsiIdensity}
\eeq
the correlator~(\ref{JJCorrelator}) is written as
\beq
\left( \Pi_q^{ {\rm bare} } (k) \right)^{ab}_{\mu \nu} = -\frac{g^2}{2} \int_q {\rm Tr}\Big[ \Gamma_\mu^{a}\calS(q_+)  \gamma_\nu^b \calS(q_-) \Big] \,,
\label{eq:pol}
\eeq
with $q_\pm = q\pm\frac{k}{2}$. The symbol ``Tr'' in Eq.~(\ref{eq:pol}) represents a trace over Dirac, color, flavor, and Nambu-Gor'kov indices. 
For the sake of clarity we write the gluon self-energy as a functional of the propagators, $\Pi=\Pi[S]$. To examine the medium effects we focus on the difference between the medium and vacuum correlators,
\beq
\left(\delta \Pi_{q}\right)_{\mu\nu}^{ab} =\left( \Pi_q^{ {\rm bare} } \left[ \calS_{ {\rm med} } \right]  \right)^{ab}_{\mu \nu} - \left( \Pi_q^{ {\rm bare} }  \left[ \calS_{ {\rm vac} } \right] \right)^{ab}_{\mu \nu}  \,,
\eeq
where $\calS_{ {\rm med} } $ and $\calS_{ {\rm vac} }$ are quark propagators in medium and in vacuum, respectively. We also note that the vertex functions are also functionals of $S$, so we write $\Gamma_\mu^a = \Gamma_\mu^a [S]$.

It is crucial to recognize the vertex functions as functionals of $S$ especially when we change the fermion bases for the loop expansion, as a naive treatment of vertices violates the conservation laws. First we eliminate problems associated with naive use of a tree level vertex so that we can focus on the regularization artifacts. For this purpose we use the WT-identity (for its derivation, see e.g. the appendix in Ref.\cite{Kojo:2014vja}),
\beq
\left\lan  D^{ac}_\mu j^c_\mu (x)  
\Psi(z_1) \bar{\Psi}(z_2) \right\ran
= - \delta^D (x-z_1) \left\lan  R_a \Psi(z_1) \bar{\Psi}(z_2) \right\ran  \nonumber\\
+ \delta^D (x-z_2)  \left\lan \Psi(z_1) \bar{\Psi}(z_2) R_a \right\ran. 
\nonumber \\
\label{WTI}
\eeq
Now we define
\beq
&\left\lan\, \left( f_{abc} A_\mu^b j_\mu^c (x) \right)\, 
\Psi(z_1) \bar{\Psi}(z_2) \right\ran \nonumber\\
&\equiv \int_{w,u} \calS(z_1-w)\, L_{\cal{NA}}^a(w-x, x-u) \, \calS (u-z_2) \,, 
\label{identity}
\eeq
where $L_{\cal{NA}}^a$ (see Fig.\ref{fig:VertexNA} for its diagrammatic structure) comes from composite operator, $\sim A j$, specific to the non-Abelian theories (the matrix $L_{\cal{NA}}^a$ appears through the WT identity but not through the perturbative expansion).

By substituting Eqs.~(\ref{JPsiIdensity}) and~(\ref{identity}) into the LHS of Eq.~(\ref{WTI}), and taking the Fourier transform and multiplying $\calS^{-1}(q_+)$ and $\calS^{-1}(q_-)$, we get
\beq
\rmi k_\mu  \Gamma_\mu^a = \calS^{-1} (q_-) R^a -  R^a \calS^{-1} (q_+) + L_{ {\cal NA}}^a \,.
\label{eq:nonabel}
\eeq
The vertex in the LHS enters our one-loop polarization function. The last term is a composite operator which is specific to non-Abelian 
theories and already contain at least 1-loop, and hence it appears only beyond the 1-loop polarization function (Fig.\ref{fig:VertexNA}). With this composite operator the WT identity is not as useful as in the QED case. However our main concern here is to illustrate the regularization artifacts associated with changes in fermion bases, and the expression is sufficient for our purpose, as we will see below.

Now we contract Eq.(\ref{eq:nonabel}) with two propagators and a tree vertex, and get the constraint for the correlator between quark color currents,
\beq
\rmi k_\mu \left( \Pi_{q }^{ {\rm bare} } \right)^{ab}_{\mu \nu} 
& = -\frac{\delta^{ab} g^2 }{4} \int_q {\rm tr}_{D,G} \Big[ \left( \calS^D(q_+) - \calS^D (q_-) \right) \gamma_\nu  \Big]  \nonumber \\
& - \frac{g^2 }{2} \int_q {\rm tr}_{c, f, D,G}  \Big[ \calS (q_+) L_{ {\cal NA}}^a \calS (q_-)  \gamma^b_\nu  \Big] \,, \nonumber \\
\label{eq:correlator_WTI}
\eeq
where in the RHS we have carried out the trace over colors and flavors for the first term, while in the second term the trace for the color, flavor, Dirac, and Nambu-Gor'kov space is all left.

As stated above, in this paper we study the gluon self-energy at one-loop. At this level of computations, non-Abelian contributions,  which at least include two loops, do not enter, and hence
\beq
\rmi k_\mu \left( \Pi_{q }^{ {\rm bare} } \right)^{ {\rm 1loop} }_{\mu \nu} 
= -\frac{g^2 }{4} \int_q {\rm tr}_{D,G} \Big[ \left( \calS^D(q_+) - \calS^D (q_-) \right) \gamma_\nu  \Big] \,, \nonumber \\
\label{eq:WT_dif}
\eeq
where 
the first and second terms in the trace should cancel if we are allowed to integrate the momentum $q$ from $-\infty$ to $+\infty$. This is the case for the dimensional regularization with which we arrive at $ k_\mu \left( \Pi_{q  }^{ {\rm bare} } \right)^{ {\rm 1loop} }_{\mu \nu} =0$.
But for the three momentum regularization, there remains a finite term as a regularization artifact. To see it, first we define 
\beq
F_\nu \left[ \calS; \vq \right] = \int \frac{\, \rmd q_0 \,}{\, 2\pi \rmi \,} \,  {\rm tr}_{D,G} \Big[ \calS^D (q) \gamma_\nu  \Big] \,,
\eeq
then $k_\mu \left( \Pi_{q }^{ {\rm bare} } \right)^{ {\cal A} }_{\mu \nu} $ is proportional to
\beq
&&\int_{\vq}  \theta( \Lambda_{ {\rm UV} }^2 - \vq^2) \left( F_\nu [ \calS; \vq_+ ] -  F_\nu [ \calS; \vq_- ] \right) \nonumber \\
&=&\int_{\vq} \left[ \theta( \Lambda_{ {\rm UV} }^2 - \vq_-^2) -  \theta( \Lambda_{ {\rm UV} }^2 - \vq_+^2) \right] F_\nu [ \calS; \vq ]   \nonumber \\
&\simeq &\int_{\vq} \, 
2 \delta( \Lambda_{ {\rm UV} }^2 - \vq^2) \, k_j q_j q_\nu F  [ \calS; \vq^2 ]  ~~~~ (F_\nu(\vq) = q_\nu F( \vq^2) ) \nonumber \\
&=&\frac{\, k_j \delta_{\nu j} \,}{\, 6 \pi^2 \,} \,  \Lambda_{ {\rm UV} }^3 \, F [ \calS; \Lambda^2_{ {\rm UV} } ] \,,
\eeq
where in the third line we dropped off higher orders of $\vec{q}\cdot \vec{k}/\Lambda_{{\rm UV}}$ which start with two extra
powers\footnote{The Taylor expansion of the step function will generate the derivatives of the delta function which look awkward. To get more well defined expressions, one can replace the step function with a smoother analytic function which interpolates $0$ and $1$ with a finite window. In such treatment $\vk/\Lambda_{{\rm UV}} \rightarrow 0$ limit can be taken in rigorous way.}.

The expression tells us that the artifact appears only if $\nu$ is spatial, which in turn means the magnetic sector. The function can be expanded by the inverse power of the UV cutoff,
\beq
F [ \calS; \Lambda^2_{ {\rm UV} } ]  = \frac{\, C_{ {\rm univ} } \,}{\,  \Lambda_{ {\rm UV} }\,} + \frac{\, C_{ {\rm dim 2} } [\calS] \,}{\,  \Lambda_{ {\rm UV} }^3 \,} + \cdots \,,
\eeq
where the first term is universal while the rest of terms depend on the quark bases.

Now we can quantify how the regularization artifact enters in $\delta \Pi$,
\beq
k_\mu \delta \Pi_{\mu\nu}^{ab} \big|_{ {\rm 3d\, reg} } \propto \delta^{ab} 
\left( C_{ {\rm dim 2} } [\calS_{ {\rm med} }] - C_{ {\rm dim 2} } [\calS_{ {\rm vac} }] \right) \,,
\eeq
where we have neglected terms of $O(\Lambda_{ {\rm UV}}^{-2})$. 
This artifact cancels when $\calS_{ {\rm med} } = \calS_{ {\rm vac} }$, but in general such equality does not hold. (As should be clear from this derivation,  the artifact is absent when $\calS_{ {\rm med} }$ and $\calS_{ {\rm vac} }$ asymptotically coincide if damping of gaps takes place at some scale $\Lambda_{{\rm damp}}$. But in this case one must manifestly take into account the momentum dependence of 
gaps. 
This makes the improvement of the vertices more complicated and will not be attempted in this paper.) 

Here we have two competing demands. On one hand it is convenient to use the dimensional regularization to be free from the artifact, but in medium the computations are cumbersome. The three-dimensional cutoff allows more straightforward calculations but it would suffer from the artifact.
In order to utilize the advantages of both regularizations we consider the following trick. We introduce a propagator which has the same structure as the vacuum one but has the mass $\tilde{M}_q$ in place of the vacuum mass $M_q^{ {\rm vac} }$,
\beq
\tilde{\calS}_{ {\rm vac} } = \calS_{ {\rm vac} } (M_q^{ {\rm vac} } \rightarrow \tilde{M}_q ) \,.
\eeq
We regroup the calculation of $\delta \Pi$ as (we suppress the Dirac and color indices for the moment)
\beq
 \delta \Pi = \delta_{\Delta \mu_q} \Pi + \delta_{\Delta S} \Pi \,,
\eeq
where
\beq
 \delta_{\Delta \mu_q} \Pi &=&  \Pi_q^{ {\rm bare} } [ \calS_{ {\rm med} } ]  -  \Pi_q^{ {\rm bare} }  [ \tilde{\calS}_{ {\rm vac} } ] \,, \\
  \delta_{\Delta S} \Pi &=&  \Pi_q^{ {\rm bare} } [ \tilde{ \calS }_{ {\rm vac} } ]  -  \Pi_q^{ {\rm bare} }  [ \calS_{ {\rm vac} }] \,.
\eeq
As for $ \delta_{\Delta S} \Pi$, there is no technical difficulty to use the dimensional regularization and {\it each} term in $ \delta_{\Delta S} \Pi$ is separately independent of artifacts. For $\delta_{\Delta \mu_q} \Pi $, the computations based on the dimensional regularization is practically not so useful. But if we choose $\tilde{M}_q$ to be $\tilde{M}^*_q$ such that
\beq
C_{ {\rm dim 2} } [\calS_{ {\rm med} }] = C_{ {\rm dim 2} } [ \tilde{\calS}_{ {\rm vac} }] \,,
\label{eq:condition_C}
\eeq
then the artifacts in $\delta_{\Delta \mu_q} \Pi $ cancel in the three-dimensional cutoff scheme. As a consequence the dimensional regularization and three-dimensional cutoff regularization become equivalent (neglecting terms of $O(\Lambda_{ {\rm UV}}^{-2})$)
\beq  
  \delta_{\Delta \mu_q} \Pi \big|_{ {\rm dim\, reg} } =  \delta_{\Delta \mu_q} \Pi \big|_{ {\rm 3d\, reg} } \,.~~~({\rm for}~\tilde{M}_q =\tilde{M}^*_q)
\eeq
Now the RHS can be computed in the standard way, without suffering from the artifact. In the next section we will perform the one-loop computation based on these regularization method.

Finally we mention that the present regularization resolves the problem of the spurious Meissner mass of two-flavor matter found in Ref.\cite{Rischke:2000qz}. Such Meissner mass was rejected in Ref.\cite{Alford:2005qw} by some subtraction method, but its justification was not claimed. The present discussion gives its justification: the subtraction method in Ref.\cite{Alford:2005qw} naturally follows from the demand to maintain the conservation law or the WT identity. Moreover the discussion here explains how to generalize the method for general sets of $M_q^{ {\rm vac}}, M_q$, and $\Delta$.

\section{One-loop result}
\label{sec:OneLoop}

Now we examine the gluon self-energy at one-loop.
As we have discussed, in general we need to use the improved vertices when we change the bases for the loop expansion. The improved vertex for one loop polarization graphs should satisfy 
\beq
\rmi k_\mu\Gamma_{\mu}^{ a}(q_-,q_+) \big|_{1{\rm loop}} = \calS^{-1}(q_-)R^a-R^a \calS^{-1}(q_+) \,,
\eeq
where the non-Abelian part appears beyond one-loop and is neglected here.
In this work we consider only the momentum independent gaps. Then it simply takes the tree level form,
\begin{eqnarray}
k_\mu \Gamma_{\mu}^{a } (q_-,q_+) = k_\mu \gamma_\mu^a \,. ~~~~~({\rm for\, 1\, loop})
\label{WTVertex} 
\end{eqnarray}
Hence we can compute Eq.(\ref{eq:pol}) with replacement $\Gamma_{\mu}^{a } \rightarrow \gamma_\mu^a$, and this vertex is sufficient to satisfy the WT identity. After handling the regularization artifacts we have $k_\mu \Pi_{\mu \nu} |_{1 {\rm loop}} =0$.

In the following we first examine the vacuum gluon self-energy without quarks, and then add quarks next. After calibrating the parameters in the theory to reproduce the lattice results, we then use them to calculate the gluon self-energy in a medium. For all comparisons from YM theory to the QC$_2$D in medium, we use the lattice data of Ref.\cite{Boz:2018crd}.

\subsection{The vacuum part}
\label{sec:vac_part}

For the comparison to the unrenormalized lattice data \cite{Boz:2018crd} we need to multiply an overall constant $Z^{ {\rm overall} }_g$. This factor should be distinguished from the conventional renormalization constant $Z_g$ that explains the difference of fields at different scales, e.g., the renormalized fields at $\mu_R$ and the bare fields at $\Lambda_{ {\rm UV}}$ as $A_{\rm bare} = Z^{1/2}_g A_R$. In contrast to $Z_g$, the overall factor $Z^{ {\rm overall} }_g$ is common for the renormalized and bare fields ($A_{\rm bare} = (Z_g^{\rm overall} )^{1/2} A'_{\rm bare}$ and $A_R = ( Z_g^{\rm overall} )^{1/2} A'_R$), and hence has nothing to do with the quantum corrections; the factor $Z^{ {\rm overall} }_g$ appear in propagators as well as the vertices, and they cancel in the final expression 
of physical quantities which should be independent of our choice of the overall normalization. 
We use the expression of a gluon propagator in vacuum with the one-loop correction  ($D^{ {\rm YM, vac} }_{\mu\nu} = D^{ {\rm YM, vac} } P_{\mu\nu} $ and $K^2 = - k^2 \ge 0$)
\beq
D^{ {\rm YM, vac} } (k)  =  \frac{Z^{ {\rm overall} }_g}{\, K^2 + m_g^2 + \Pi_{ {\rm YM, vac}} (k) \,} \,.
\eeq
As described in Eq.(\ref{SU2GluonRC}) our renormalization conditions are $\Pi_{ {\rm YM, vac}} (\mu_R^2)  = \Pi_{ {\rm YM, vac}} (0)  = 0$ \cite{Tissier:2011ey} which determine our counter terms $\delta_{Z}$ and $\delta m_g^2$. Meanwhile the constant $Z^{ {\rm overall} }_g$ is adjusted as
\beq
D^{ {\rm YM, vac} } (\mu_R) = \frac{ Z^{\rm overall}_g }{\, \mu_R^2 + m_g^2 \,} = D^{ {\rm YM, vac} }_{ {\rm lat}} (\mu_R) \,.
\eeq
 Because our theory is chosen for the description at low energy, we take $\mu_R \sim 1$ GeV. 

We will not put too much effort for the precise fit beyond $\sim 1$ GeV because some discrepancy is expected due to the use of constant $m_g$, which actually should be momentum dependent and be vanishing in the UV limit, and also due to the lack of the RG improvement in the present work. Hence, if one compares the one-loop results for $K^2 D(k)$ to the lattice's, the large momentum behaviors look different for the reasons rather obvious to us. We postpone the calibration of the UV part to the future studies and focus more on the behaviors up to $K \sim 1$ GeV.

\subsubsection{The YM part}
\label{sec:YM_part}

%
\begin{table}[t]
\vspace{0.5cm}
\hspace{2.5cm}
\begin{tabular}{|r|c|c||c|c|c|}
\hline 
~~~~~~&~$L\, [{\rm fm}]~$~&~$a^{-1}$[GeV]~~&~$m_g$[GeV]~~&~~~$Z_g^{ {\rm overall} }$~~~\\ \hline 
~YM 1~~&~6.5~&~0.91~&~0.68~&~4.0~~\\
      ~2~~&~4.3~&~1.59~&~0.66~&~4.5~~\\ \hline
     
~vac A~~&~8.5~&~0.74~&~0.85~&~2.8~~\\
        B~~&~7.4~&~0.86~&~0.85~&~2.8~~\\ 
        C~~&~6.0~&~1.10~&~0.74~&~3.2~~\\
        D~~&~2.2~&~1.40~&~0.48~&~4.5~~\\ \hline
\end{tabular}
\caption{\footnotesize{The parameters used for the fit. 
The first two columns are parameters in the lattice simulations; $L$ and $a$ are the box size and lattice spacing. The last two columns are parameters used in our model calculations. For all fits we used $\alpha_s=1$-$3$ whose variation is reflected in the error band.}
 }
\label{tab:vac}
\end{table}
%

\begin{figure}[thbp]
\centering
\includegraphics[scale=0.7]{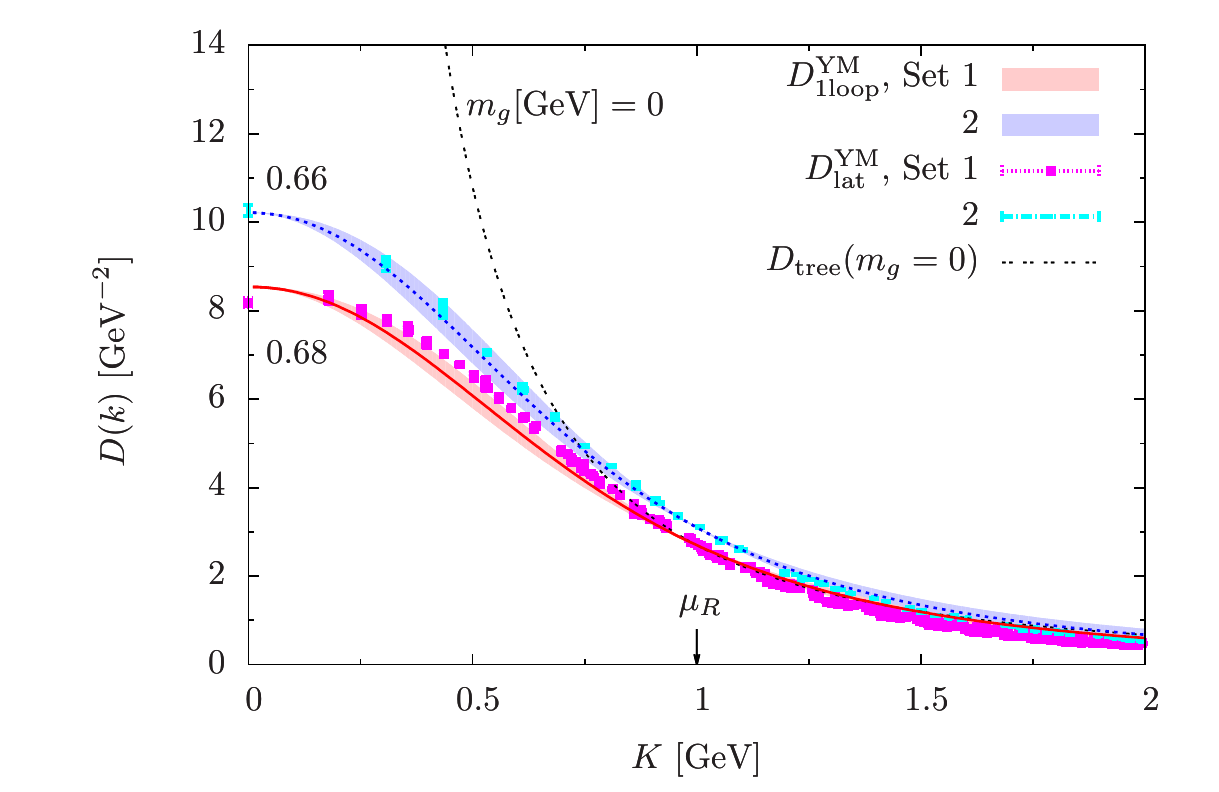}
 \caption{(color online) The gluon propagators in the YM theory renormalized at $\mu_R=1$ GeV. The dots indicate the lattice results~\cite{Boz:2018crd} for the sets YM 1 and 2 shown in Table.\ref{tab:vac}, compared to the one-loop calculations with a fitting parameter $m_g$. We put the error band which comes from the variation, $\alpha_s =1$-$3$, in the one-loop result. The line inside of the band corresponds to the $\alpha_s=2$ results. With larger $\alpha_s$, the one-loop result of $D_g$ below (above) 1 GeV is more enhanced (reduced).
 }
\label{fig:YM_Dg_mg}
\end{figure}

We first discuss the YM case. The contribution from the massive pure YM theory can be written as \cite{Tissier:2011ey}
\begin{eqnarray}
\Pi_{ {\rm YM} }(k) 
&=& \frac{g^2 K^2}{192\pi^2}\Bigg\{111{s}^{-1}-2{s}^{-2}+(2-{s}^2){\rm ln}({s}) \nonumber\\
&+& 2({s}^{-1}+1)^3({s}^2-10{s}+1){\rm ln}(1+{s}) \nonumber\\
&+& (4{s}^{-1}+1)^{3/2}({s}^2-20{s}+12) \nonumber\\
&\times & {\rm ln}\left(\frac{\sqrt{4+{s}}-\sqrt{{s}}}{\sqrt{4+{s}}+\sqrt{{s}}}\right)-({s} \leftrightarrow {\mu}_R^2/m_g^2)\Bigg\} \ , 
\label{PiGc} 
\end{eqnarray}
with $s = K^2/m_g^2$. Below we examine the values of the coupling constant and the gluon mass that can fit gluon propagators in the lattice results.

The lattice results \cite{Boz:2018crd} for gluon propagators in the infrared are sensitive to the finite volume effects (for systematic studies, see Ref.\cite{Fischer:2007pf,Bornyakov:2009ug}) and we need to decide which data to be fitted. 
The general trend is that with larger volume the gluon propagators are more suppressed in the infrared. Without taking the volume sufficiently large, we tend to underestimate the size of $m_g$. 
For illustration purposes we plot the largest and second largest volume results in the lattice data of \cite{Boz:2018crd} to show the impact of finite size effects on the estimate of $m_g$.

Shown in Fig.\ref{fig:YM_Dg_mg} is the comparison of the one-loop and the lattice results for gluon propagators with the sets YM1 and YM2 listed in Table.\ref{tab:vac}.  When we fit each lattice data set, we first adjust the overall normalization $Z_g^{ {\rm overall} }$ at $\mu_R=1$ GeV, and then search the non-perturbative parameter $m_g$ which gives the good fit. The value of $\alpha_s$ is changed from 1 to 3 and we attached the error band around the line given at $\alpha_s=2$. The infrared behavior is most sensitive to the choice of $m_g$, and we found that $m_g \simeq 0.66$-$0.68$ GeV fits the data well. Meanwhile the loop corrections (which are regular in the infrared) damp as $\sim K^2$ at small momenta and hence the details do not have much impacts in the deep infrared. In fact the variation of $\alpha_s$ from 1 to 3 changes the gluon propagators only modestly in the infrared. Because we renormalize the self-energy as $\Pi(\mu_R=1\,{\rm GeV})=0$, the different propagators coincide at $K =1$ GeV. Below this momentum the loop corrections enhance the propagator while suppress it at large momenta.

\subsubsection{The vacuum part with quarks}
\label{sec:vac_part_with_quarks}

\begin{figure}[thbp]
\centering
\includegraphics[scale=0.7]{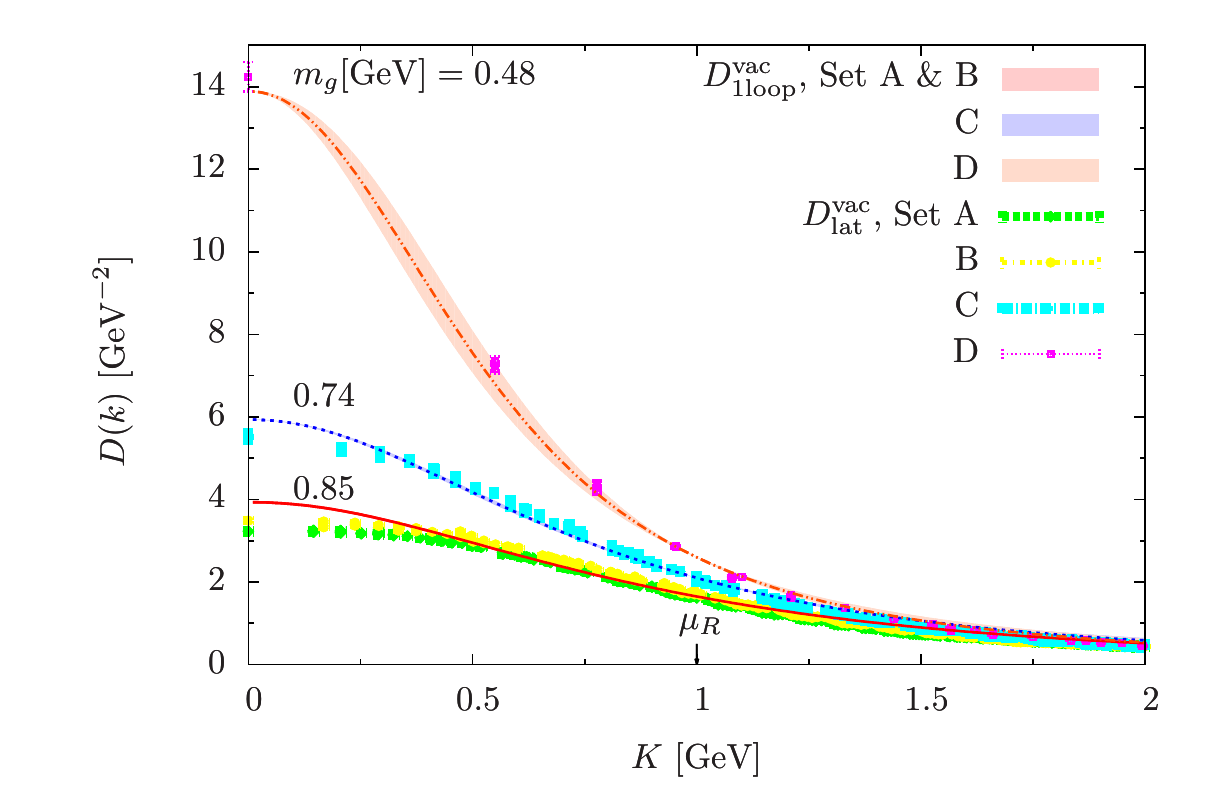}
 \caption{(color online) The gluon propagators in vacuum with dynamical quarks, renormalized at $\mu_R=1$ GeV. The dots indicate the lattice results~\cite{Boz:2018crd} for the sets vac A-D shown in Table.\ref{tab:vac}, compared to the one-loop calculations. The finite volume effects are large and accordingly the values of $m_g$ used for the fits vary considerably. The error bands for the sets A-C with large $m_g$ are tiny and not visible. }
\label{fig:vac_Dg}
\end{figure}

Next we include quarks. The quark contribution is
\beq
&&\Pi_q^{ {\rm vac}} (k)  = - K^2\frac{g^2}{2\pi^2} \nonumber \\
&& \times \int_0^1 dx\, x(1-x) {\rm ln} \frac{ (M^{ {\rm vac} }_q)^2 + x(1-x) K^2 }{\, (M^{ {\rm vac} }_q)^2 + x(1-x) \mu_R^2 \,}\ .
\eeq
There are four lattice data sets with different volumes \cite{Boz:2018crd}.  The values used for the fit are listed in Table.\ref{tab:vac} as the vac A-D. As in the YM case we choose $\alpha_s = 1$-$3$ and associate the error band. We also checked how the results depend on the effective quark mass $M_q^{{\rm vac}}$, and found that its impacts are negligible for $M_q^{{\rm vac}} =0.1$-$0.3$ GeV. Below we use $M_q^{{\rm vac}} =0.3$ GeV.

Shown in Fig.\ref{fig:vac_Dg} are the comparison between the lattice data and the one-loop results. The finite volume effects in the lattice data are very large in the infrared and accordingly our choices for $m_g$ vary considerably, from $m_g =0.48$ to $0.85$ GeV. The lattice results for the largest volume favors $m_g \simeq 0.85$ GeV. At this point we are not very sure about the value of $m_g$ and further examination of finite volume effects as well as the discretization artifacts is called for. Nevertheless, it seems safe to conclude that the gluon propagators with larger $m_g$ become more insensitive to the value of $\alpha_s$ in the infrared. 

This indicates that, by including strong coupling effects into the effective residue and mass in gluon propagators, the residual strong coupling effects may be treated as small corrections. This is the key feature for quasi-particle descriptions.

\begin{figure}[thbp]
\centering
\includegraphics[scale=0.7]{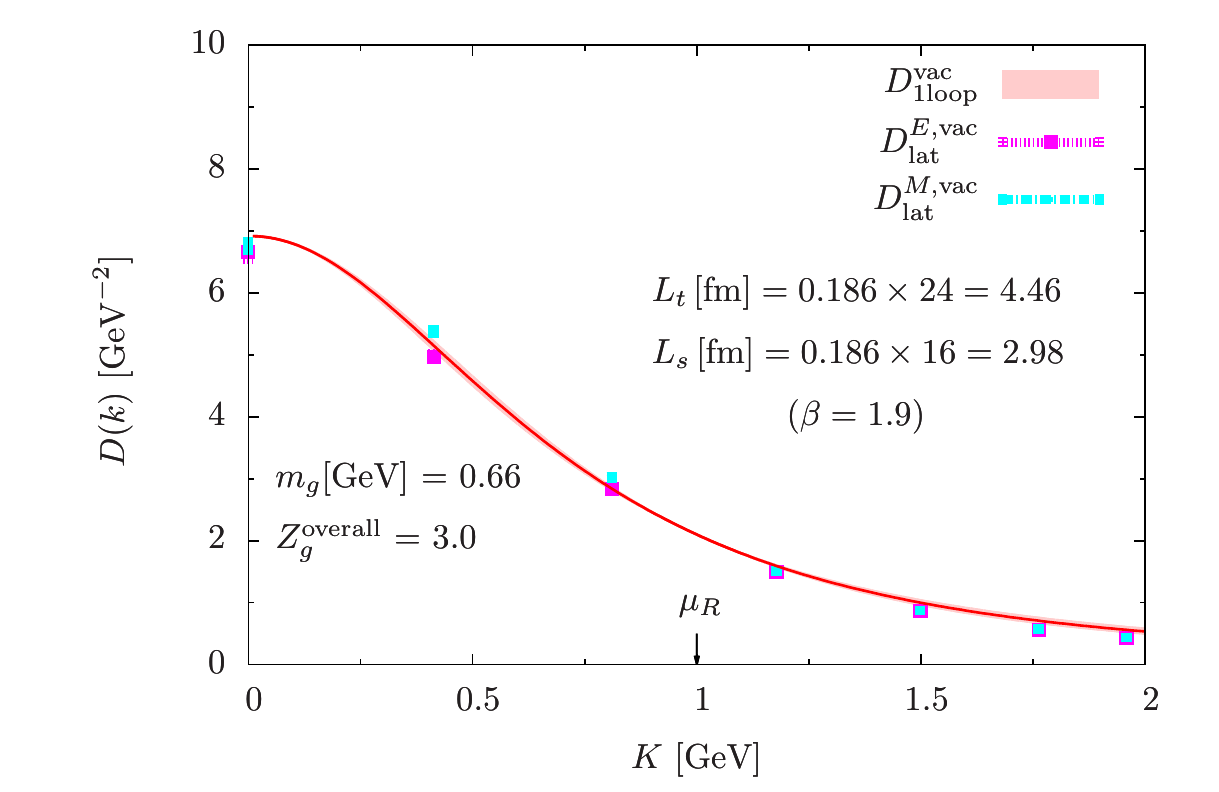}
 \caption{(color online) The gluon propagators in vacuum with dynamical quarks, renormalized at $\mu_R=1$ GeV. The setup for this lattice data will be also used for the medium propagator. In spite of the vacuum results, the lattice's electric and magnetic propagators slightly differ due to the artifacts of using the anisotropic lattice. }
\label{fig:Dg_med_mu318}
\end{figure}

Having examined the finite volume effects, below we fix our parameters to fit the lattice data for $\beta = 1.9$ and $N_t \times N_s^3 = 24 \times 16^3$, although this is not the best quality in the available data. The reason to choose this set is that it was used for the lattice simulations in medium, see the next section. In medium computations the size in the temporal and spatial directions are often taken to be different,
\beq
L_t  &=& a N_t = 0.186\, {\rm fm} \times 24 = 4.46\, {\rm fm} \,, \nonumber \\
L_s &=& a N_s = 0.186\, {\rm fm} \times 16 = 2.98\, {\rm fm} \,,
\eeq
and hence the propagators for electric and magnetic gluons may differ even 
in vacuum\footnote{Strictly speaking, in \cite{Boz:2018crd} no lattice data are available for $\mu_q=0$ in this setup. But there is data at $\mu_q=318$ MeV below the matter threshold, $\mu_{c} = m_\pi/2$ is $\sim 380$ MeV for the heavy pion mass used in this simulation. Thus we can regard the result at $\mu_q=318$ MeV as the vacuum result. }. 
For this set the electric and magnetic propagators do not differ 
much\footnote{In \cite{Boz:2018crd} there are other set of lattice data, $\beta = 2.1$ and $N_t \times N_s^3 = 32 \times 16^3$, but the volume is smaller, $L_t=4.4$ fm and $L_s=2.2$ fm. The artifacts of anisotropy is much stronger than the case we are studying. So we omit this case from our study.}.
Taking the renormalization scale to be $\mu_R = 1$ GeV as before and adjusting the overall normalization to be $Z_g^{ {\rm overall} }=3.0$, our one-loop propagator fits the data well for $m_g = 0.66\, {\rm GeV}$ for $\alpha_s = 1$-$3$. The quality of the fit can be seen in Fig.\ref{fig:Dg_med_mu318}. 

The comparison between Figs.\ref{fig:vac_Dg} and \ref{fig:Dg_med_mu318} seems to suggest that the volume used for the medium is not small enough for decisive statements and the value of $m_g$ tends to be underestimated. Keeping this in mind, in what follows we take $m_g = 0.66\, {\rm GeV}$ as a reference point to examine the medium effects for the set $\beta = 1.9$ and $N_t \times N_s^3 = 24 \times 16^3$.

\subsection{The medium part}
\label{sec:medium_part}

\subsubsection{The determination of $\tilde{M}_q$ }
\label{sec:tildeM}

Now we turn to the gluon propagator in the medium. As we have detailed in Sec.\ref{sec:regularization}, we must handle the regularization artifact. The medium computations in our regularization involves the parameter $\tilde{M}_q$ and it is fixed according to the condition Eq.(\ref{eq:condition_C}). To examine this condition we calculate the function $F_\nu$ explicitly,
\beq
F_\nu \left[ \calS; \vq \right] 
&=& \int \frac{\, \rmd q_0 \,}{\, 2\pi \rmi \,} \,  {\rm tr}_{D} \Big[ \left( \calS_{11}^D (q) + \calS_{22}^D (q) \right) \gamma_\nu  \Big] \nonumber \\
&=& 2 \delta_{\nu i} q_i  \sum_{ s = \rp, \ra } \frac{\, | u_s^2(\vq)|^2 -  | v_s^2(\vq)|^2 \,}{E_q}
\,.
\eeq
At large $|\vq|$, 
\beq
 \frac{\, | u_s^2(\vq)|^2 -  | v_s^2(\vq)|^2 \,}{E_q}~\sim~  \frac{1}{\, |\vq| \,}  - \frac{\, \Delta^2 + M_q^2 \,}{\, 2 |\vq|^3 \,} + \cdots \,.
\eeq
\\
The expression for $F_\nu [ \tilde{\calS}_{{\rm vac}}; \vq ] $ is obtained by replacement, $\mu_q, \Delta \rightarrow 0$ and $M_q \rightarrow \tilde{M}_q$.
In order to achieve the condition $C_{ {\rm dim 2} } [\calS_{ {\rm med} }] = C_{ {\rm dim 2} } [ \tilde{\calS}_{ {\rm vac} }]$, we take
\beq
\tilde{M}_q = \sqrt{\,   \Delta^2 + M_q^2 \,} \,,
\eeq
with which the medium calculations in three dimensional cutoff are free from the UV artifacts.

Now we compute the medium self-energy $\Pi = \Pi_{ {\rm vac} } + \delta \Pi$ where $ \delta \Pi = \delta_{\Delta S} \Pi + \delta_{\Delta \mu_q} \Pi $.

\subsubsection{The computation of  $\delta_{\Delta S} \Pi $ }
\label{sec:delta_SPi}

The function $\delta_{\Delta S} \Pi^{} (k) = \Pi_q^{ {\rm bare} } [ \tilde{ \calS }_{ {\rm vac} } ]  -  \Pi_q^{ {\rm bare} }  [ \calS_{ {\rm vac} }]$ measures the modification associated with the changes in vacuum bases for quark propagators. It can be  computed in the dimensional regularization as
\beq
&& \delta_{\Delta S} \Pi^{} (k) \big|_{ {\rm dim\, reg} }  = - K^2\frac{g^2}{2\pi^2} \nonumber \\
&& \times \int_0^1 dx\, x(1-x) {\rm ln} \frac{ ( \tilde{M}_q )^2 + x(1-x) K^2 }{\, (M^{ {\rm vac} }_q)^2 + x(1-x) K^2 \,}\, ,
\eeq
which enters the electric and magnetic components in the same way. This contribution approaches zero for vanishing momenta.

\subsubsection{The computation of  $\delta_{\Delta \mu_q} \Pi $ }
\label{sec:delta_muPi}

Next we present the results of $\delta_{\Delta \mu_q} \Pi $. There are three distinct medium contributions; the particle-hole (pp), antiparticle-antihole (aa), and particle-antiparticle (pa) contributions. The electric and magnetic parts of the quark one-loop self-energy are calculated in the three dimensional regularization as in Ref.\cite{Kojo:2014vja},
\begin{widetext}
\begin{figure}[thbp]
\centering
\includegraphics[scale=0.68]{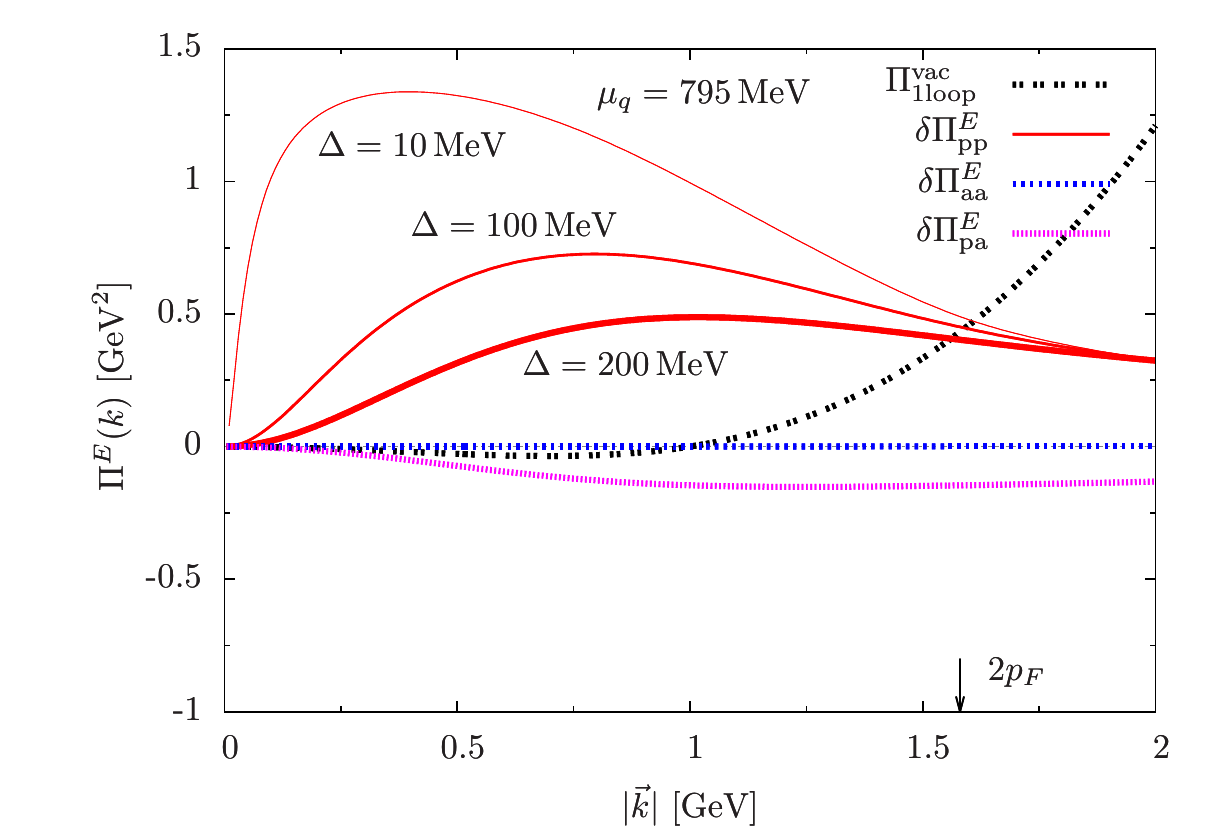}
 \caption{(color online) The polarization in the electric channel for $\mu_q=795\, {\rm MeV}$, $\alpha_s=2$, $\Delta=200\,{\rm MeV}$, and $M_q=100$ MeV. The particle-hole contributions are very sensitive to the value of $\Delta$, so only in this channel we also plot the results of $\Delta=10$ and $100$ MeV. The vacuum polarization renormalized at $\mu_R=1$ GeV is also plotted as a reference. The allow indicates $2p_F \simeq 1.58$ GeV.
 }
\label{fig:PiE_med_mu}
\end{figure}
\begin{figure}[thbp]
\centering
\includegraphics[scale=0.68]{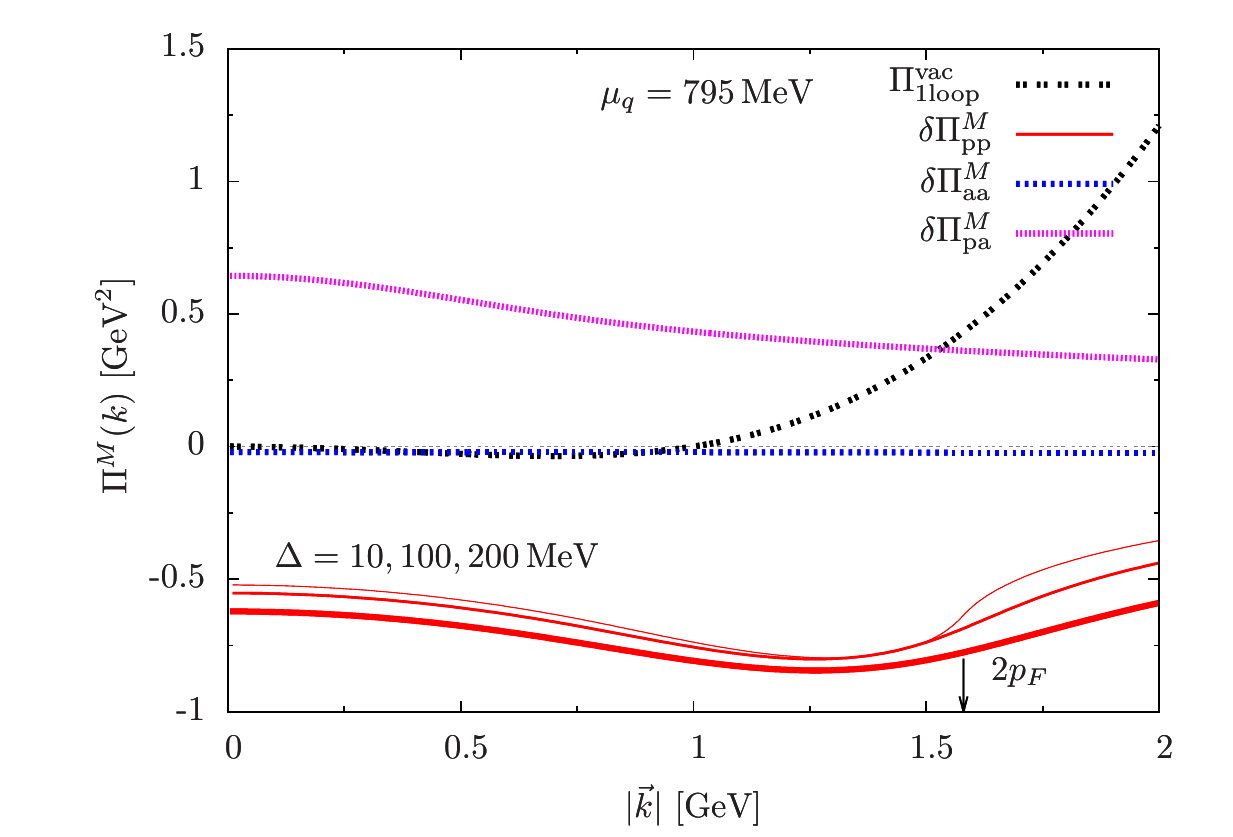}
 \caption{(color online) The polarization in the magnetic channel plotted in the same way as Fig.\ref{fig:PiE_med_mu}. The particle-hole contribution has a negative peak near $2p_F$ and approaches zero at higher momenta. The impacts of variation $\Delta=10$-$200$ MeV are small in all channels.
  }
\label{fig:PiM_med_mu}
\end{figure}
%
\begin{eqnarray}
\delta_{ \Delta \mu_q} \Pi^q_{E,M}(k) \big|_{ {\rm 3d\, reg} } 
= g^2  \sum_{ s, s' = \rp, \ra } \int_{ \vec{q} } \left[\,  {\cal C}_{E,M}^{ss'} ( \vq_+, \vq_- ) \,  {\cal K}_{E,M}^{ss'} (\vq_+, \vq_-) \, {\cal G}_{ss'} (q_+,q_-) - (M_q \rightarrow \tilde{M}_q;~~ \mu_q, \Delta \rightarrow 0) \,\right] \,.
\end{eqnarray}
\end{widetext}
Here ${\cal C}$'s and ${\cal K}$'s are the coherence and kinematic factors, respectively, which differ for magnetic and electric polarizations. The former is sensitive to the quantum numbers (color, flavor, spin) of condensates which decide whether the normal and anomalous contributions add coherently or incoherently. Meanwhile ${\cal K}$'s reflect the kinematical structure of the spinor bi-linears. These factors depend only on the static momenta. The factors ${\cal G}$ are the propagators which reflect the pole structures. Only this part depends on $k_4$ and hence is totally responsible for the dynamical aspect of the gluon polarization. 
\begin{table}[bht]
\vspace{0.5cm}
\hspace{2.5cm}
\begin{tabular}{|r| c|c || c|c || c|}
\hline 
~~~~~~&~${\cal C}_E$~&~${\cal K}_E$~&~${\cal C}_M$~&~${\cal K}_M$~&~${\cal G} ( |\vq|=p_F)$~ \\ 
\hline 
~~pp~~~&~$\sim \vk^2$~&~2~&~$\left( \frac{\Delta}{\epsilon_{ {\rm p} } (q) } \right)^2$~&~$\left( \frac{ \vq}{E_q} \sin \theta \right)^2$~&$\sim \frac{1}{\, \vk^2 + \Delta^2 \,}$ \\
~~aa~~~&~$\sim \vk^2$~&~2~&~$\left( \frac{\Delta}{\epsilon_{ {\rm a} } (q) } \right)^2$~&~$\left( \frac{ \vq}{E_q} \sin \theta \right)^2$~&$\sim \frac{1}{\, p_F \,}$  \\
~~pa~~~&~finite~~&~$ \sim \vk^2$~~&~finite~~&~~$-2$~~&$\sim \frac{1}{\, p_F \,}$  \\ 
\hline
\end{tabular}
\caption{\footnotesize{ The coherence and kinematical factors for electric and magnetic gluons, and the factors from the propagators at $|\vq|=p_F$ where $p_F$ is the quark Fermi momentum such that $E(p_F)=\mu_q$.
 } }
\label{tab:medium}
\end{table}
%
\begin{figure*}[htbp]
\centering
\includegraphics*[scale=0.52]{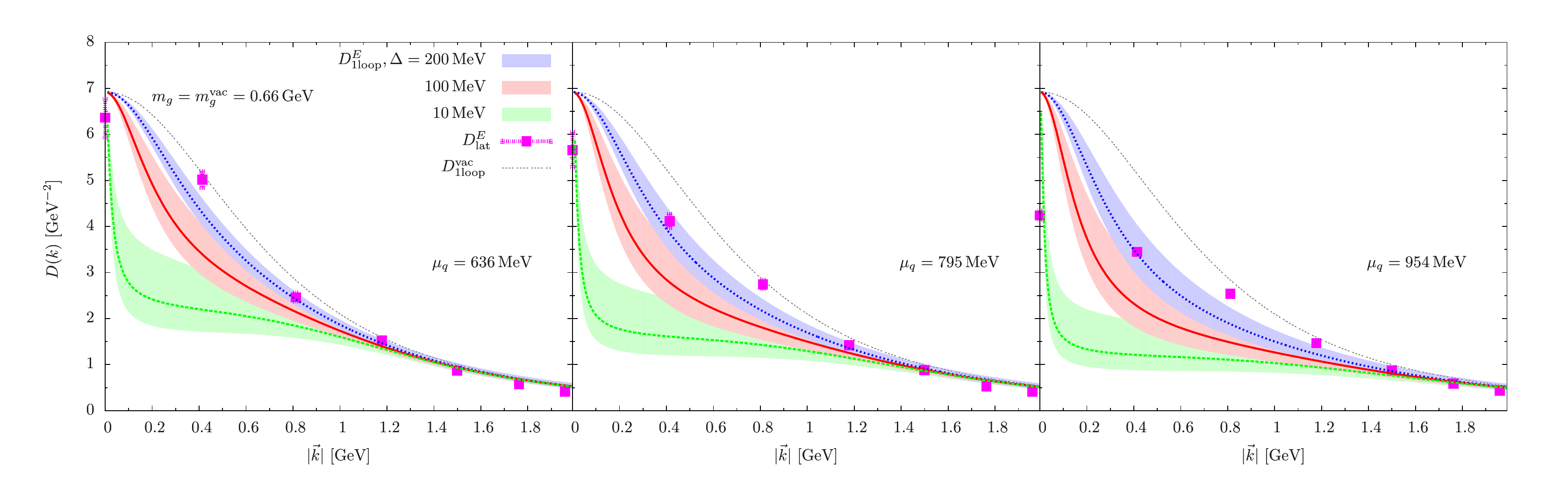}
 \caption{(color online) The in-medium electric gluon propagators at one-loop for $\alpha_s=1$-$3$ and the gaps $\Delta =10, 100$, and $200$ MeV. Larger $\alpha_s$ reduces more the propagator at finite momenta ($ |\vk| \gtrsim \Delta$) by the screening effects. The discrete data points are from the lattice results. The vacuum result is also shown for a reference.}
\label{fig:Dg_E_dmu}
\end{figure*}

\begin{figure*}[htbp]
\centering
\includegraphics*[scale=0.52]{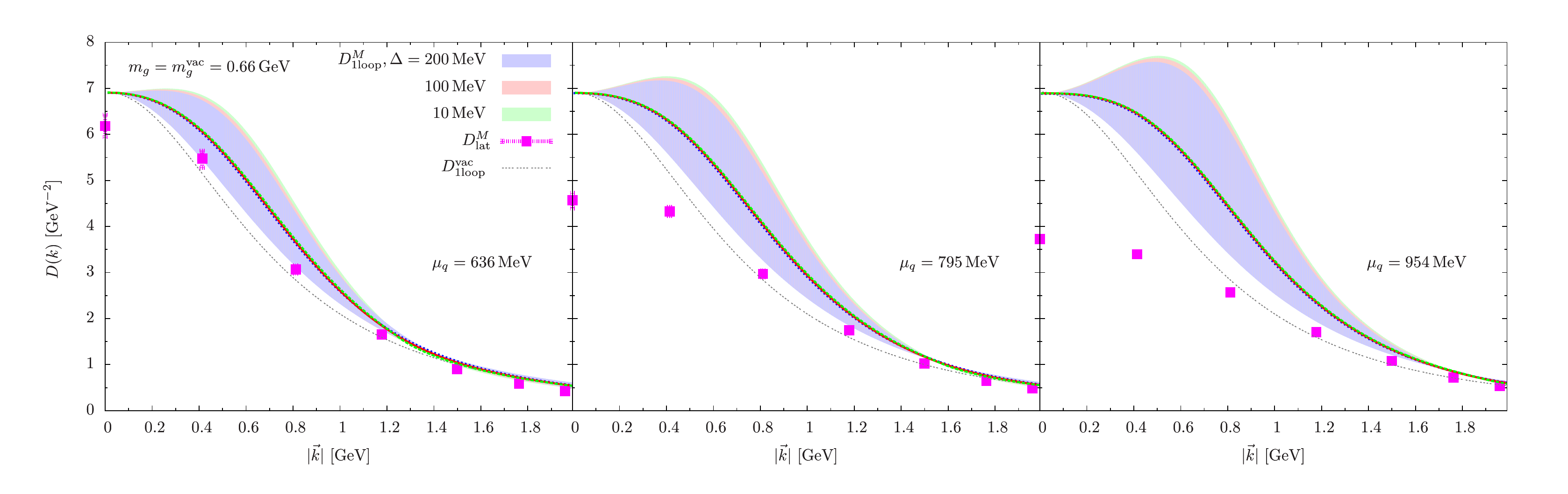}
 \caption{(color online) The magnetic gluon propagators for the setup same as Fig.\ref{fig:Dg_E_dmu}. With larger $\alpha_s$, the polarization effects enhance more the propagators at finite momenta (the {\it para}-magnetic contributions dominate over the {\it dia}-magnetic ones). The difference in $\Delta$ has little impact and is difficult to see in this plot.}
\label{fig:Dg_M_dmu}
\end{figure*}

The explicit forms of these factors are as follows: the coherence factors are
\begin{eqnarray}
{\cal C}^{ \rp \rp}_{E,M} &=& \frac{1}{2} \left( 1 - \frac{\, ( E_{q_+} - \mu_q  )(  E_{q_-}  - \mu_q ) \pm |\Delta|^2 \,}{ \epsilon_\rp (q_+) \epsilon_\rp (q_-) } \right) \,, \nonumber\\
{\cal C}^{ \ra \ra}_{E,M} &=& \frac{1}{2} \left( 1 - \frac{\, ( E_{q_+} +  \mu_q )( E_{q_-} + \mu_q ) \pm |\Delta|^2 \,}{ \epsilon_\ra (q_+) \epsilon_\ra (q_-) } \right) \,, \nonumber\\
{\cal C}^{ \rp \ra}_{E,M} &=& \frac{1}{2} \left( 1 + \frac{\, ( E_{q_+} -  \mu_q )( E_{q_-} + \mu_q ) \mp |\Delta|^2 \,}{ \epsilon_\rp (q_+) \epsilon_\ra (q_-) } \right) \,, 
\end{eqnarray}
where ${\cal C}^{ \rp \ra}_{E, M} (q_+, q_-) = {\cal C}^{ \ra \rp}_{E,M} (q_-, q_+)$; the kinematic factors are
\begin{eqnarray}
&& {\cal K}^{ \rp \rp}_{E} =  {\cal K}^{ \ra \ra}_{E}  = 1 + \frac{\, \vec{q}^2 - \vec{k}^2/4 + M_q^2 \,}{ E_{q_+}E_{q_-} }  \ , 
\nonumber \\
&& {\cal K}^{ \rp \ra }_{E} =  {\cal K}^{ \ra \rp}_{E}  =  1 - \frac{\, \vec{q}^2 - \vec{k}^2/4 + M_q^2 \,}{ E_{q_+}E_{q_-} } \ , 
\nonumber \\
&& {\cal K}^{ \rp \rp}_{M} =  {\cal K}^{ \ra \ra}_{M}  = -1 + \frac{\, ( |\vec{q}|\cos\theta)^2 - \vec{k}^2/4 + M_q^2 \,}{ E_{q_+}E_{q_-}}  \ ,
\nonumber \\
&& {\cal K}^{ \rp \ra}_{M} =  {\cal K}^{ \ra \rp}_{M}  = -1 - \frac{\, ( |\vec{q}|\cos\theta)^2 - \vec{k}^2/4 + M_q^2 \,}{ E_{q_+}E_{q_-}}  \ ,
\end{eqnarray}
where $\hat{k} = \vk/|\vk|$ and $\cos\theta$ is the angle between $\vq$ and $\vk$; finally the propagator part is
\beq
{\cal G}_{ss'} (q_+,q_-) &=& \frac{1}{\, 2 \,} \bigg( \frac{1}{\, \rmi k_4 + \epsilon_s(q_+) + \epsilon_{s'} (q_-) \,} \nonumber \\
&& ~~ + \frac{1}{\, - \rmi k_4 + \epsilon_s(q_+) + \epsilon_{s'} (q_-) \,} \bigg) \,.
\eeq
Below we focus on the static behaviors of the gluon propagators at $k_4=0$.

The qualitative behaviors of these factors in the limit $k_4=0$ and $|\vk| \rightarrow 0$ are summarized in \cite{Kojo:2014vja}. For practical convenience we reproduce it in Table.\ref{tab:medium}.  

Now we examine the behavior of the in-medium contributions, $\delta \Pi^{E,M}$, from the electric and magnetic sectors in the singlet phase. We consider $\mu_q > 600$ MeV and assume that the chiral mass $M_q$ is close to the current quark mass. The lattice data to be compared (next subsection) have used the current quark mass of $m_q \sim 100$ MeV, so we fix $M_q = 100$ MeV. Through our attempts in fitting, we found that the details of $M_q$ are not important.

Figs.\ref{fig:PiE_med_mu} and \ref{fig:PiM_med_mu} show our results for the electric and magnetic polarization functions at one-loop. We examine the roles of pp-, aa-, pa-channels and their dependence on the gap. We took the parameters, $\mu_q=795$ MeV, $\alpha_s=2$, $\Delta =200$ MeV, and $M_q=100$ MeV. As the particle-hole contributions are most sensitive to the value of the gap, we also plot the results of $\Delta=10$ and $100$ MeV with thin lines.

The electric polarization function in the infrared is largely dominated by the particle-hole (pp-) contributions. In particular the in-medium electric screening mass is saturated by the particle-hole contributions near the Fermi surface. For $|\vk|\rightarrow 0$ the coherence factor $ {\cal C}^{ \rp \rp}_{E}$ vanishes. In the absence of gaps the static particle-hole propagator ${\cal G}_{\rp \rp}$ has the IR divergence of $\sim 1/\vk^2$ for $|\vk|\rightarrow 0$, so the product of $ {\cal C}^{ \rp \rp}_{E}$ and ${\cal G}_{\rp \rp}$ becomes finite yielding the Debye mass. With finite gaps the IR divergence from ${\cal G}_{\rp \rp}$ is regulated, so the product vanishes as $\sim \vk^2/\Delta^2$ for $|\vk|\rightarrow 0$, resulting the vanishing electric screening mass. The difference between the normal phase with $\Delta=0$ and the singlet phase is seen up to $\sim 2\Delta$ beyond which the effects arising from the different phase structures become negligible. 

Meanwhile the magnetic polarization function is much less affected by the details of the gaps. The absence of the magnetic screening mass is achieved only after summing up all the contributions, ($\rp\rp, \rp\ra, \ra\ra$) parts, which are related in an intricate way by the gauge invariance. Here it should be emphasized that the particle-hole and antiparticle-antihole contribute as the {\it para}-magnetic effects which {\it enhances} the propagation of magnetic gluons, while the particle-antiparticle (with the vacuum subtraction) contributes as the {\it dia}-magnetic effects that suppress magnetic gluons. For $\vk\rightarrow 0$ these contributions precisely cancel {\it if} we correctly maintain the conservation law or the WT identity \cite{Kojo:2014vja}. At finite momenta, the para-contributions win and the magnetic gluon propagators are enhanced from the vacuum one. The para-contribution is maximized around $2p_F$ and then approaches zero at higher momenta. Finally we mention that, if we treat the Higgs phase instead of the singlet (or normal) phase, the coherence factor vanishes as $\sim \vk^2$ and the para-contributions are suppressed; then the particle-antiparticle contributions dominate to screen the magnetic gluons, resulting in the Meissner mass.

\subsubsection{Comparison with the lattice data }
\label{sec:compare_lattice}

Now we examine the electric and magnetic gluon propagators by adding $ \delta \Pi = \delta_{\Delta S} \Pi + \delta_{\Delta \mu_q} \Pi $ to the vacuum polarization tensors. The gluon mass $m_g$ and overall normalization $Z_g$ are kept fixed to the vacuum one ($m_g=0.66$ GeV and $Z_g=3.0$) so that the medium dependence of the propagator should be regarded as the prediction of one-loop calculations. The results of one-loop calculations are shown in Figs.\ref{fig:Dg_E_dmu} and \ref{fig:Dg_M_dmu} together with the lattice data ($\beta = 1.9$, $N_t \times N_s^3 = 24 \times 16^3$) at $\mu_q=636, 795$, and $954$ MeV.

As for the electric sector, the comparisons seem to suggest that the phase structure or quark pairings are important for the explanation of the lattice results. With too small gaps ($\Delta \lesssim 100\,{\rm MeV}$), at $|\vk| \gtrsim 2\Delta$ the polarization function approaches the normal phase result which are dominated by the Debye screening scale of $\sim g\mu_q$, and the resulting gluon propagator is screened too much. We varied $\alpha_s$ from $1$-$3$ but such variation does not improve the situation. Therefore we first conclude that the inclusion of the gap is crucial to obtain the reasonable fit. Having concluded that, we also emphasize that the use of the gap of $\Delta =100$-$200$ MeV does not fully explain the lattice results, especially the tendency at $|\vk|$ less than $\sim 0.5$ GeV. The agreement becomes worse at larger $\mu_q$. In particular Fig.\ref{fig:Dg_E_dmu} shows that the lattice electric propagator has the larger screening mass than the vacuum one, in contrast to the one-loop prediction for the singlet phase. This discrepancy likely indicates the lack of the relevant physics in the one-loop result. For example, gluon propagators inside of loops remain the vacuum one but this is not a consistent treatment when the in-medium effects become large. We will come back to this point in Sec.\ref{sec:pheno_fit}.

The situation seems more problematic in the magnetic sector. Here the polarization functions are almost degenerate in the normal and paired phases. Indeed, changing $\Delta$ does not improve the consistency between the lattice results and the one-loop prediction. Two features are particularly noteworthy: i) the one-loop result predicts the absence of the medium induced magnetic mass, but the lattice results seem to suggest the enhancement of the magnetic mass; ii) at finite momenta the gluon propagator in the one-loop result is significantly enhanced from the vacuum one by the para-magnetic contributions associated with the particle-hole channels, but in the lattice results the changes are much more modest or absent. 

To summarize this section, we found that the inclusion of the quark gaps improves the consistency between the one-loop calculations and lattice results. Another important point is that, starting with the propagator with the gluon mass, the importance of the medium effects are tempered; as the denominator of the propagator already has some mass scales of $\sim 0.5$-$0.8$ GeV even before the medium effects enter there. Having said that, we must also conclude that the consistency at this level of analyses is not quite satisfactory especially in the magnetic sector. In fact the lattice results suggest that the electric and magnetic propagators behave similarly in contrast to the one-loop propagator. The obvious deficiency in the one-loop approach was that we kept using the vacuum gluon propagators inside of the loop in spite of the fact that the one-loop vacuum and medium propagators start to deviate already around $ |\vk| \simeq 1$ GeV. We expect the reduction in the electric propagator and the enhancement in the magnetic propagator to be somehow averaged out as the electric (magnetic) gluon propagator enters the loop for the magnetic (electric) polarization. To take into account such effects one must perform the renormalization group (RG) improvement. We postpone such analyses to our future project.

\section{Phenomenological inspection of non-perturbative effects}
\label{sec:pheno_fit}

\begin{figure*}[thbp]
\centering
\includegraphics*[scale=0.52]{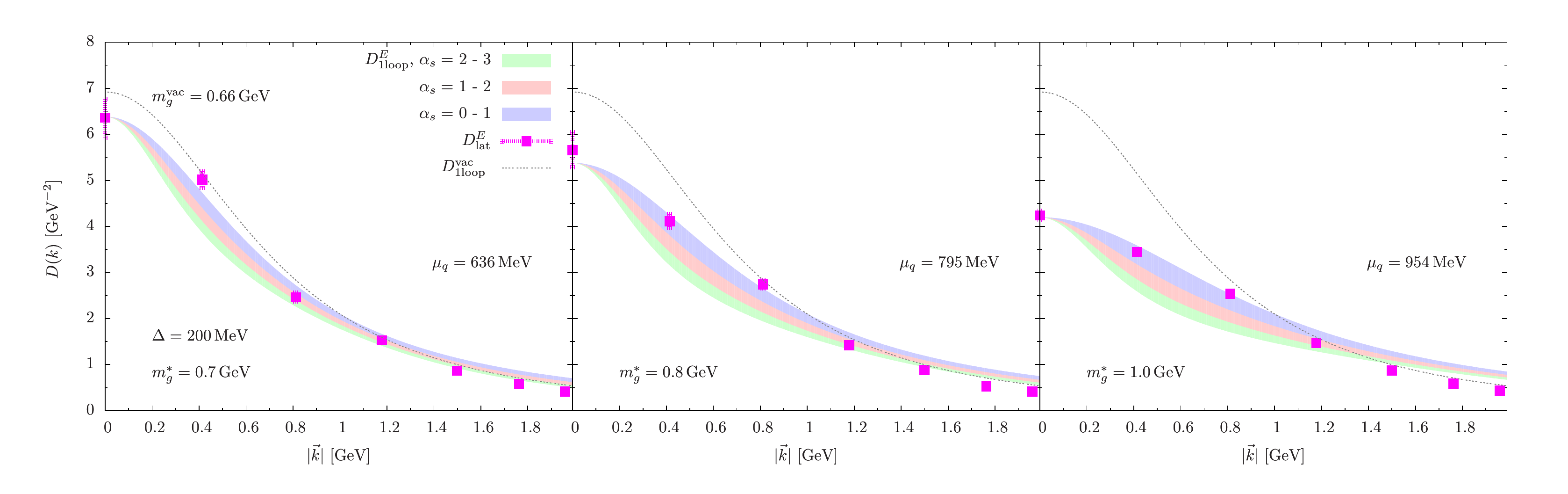}
 \caption{(color online) The electric gluon propagator at various chemical potentials. $Z_g^{\rm overall}$ is kept fixed to the vacuum value and we chose $\Delta =200$ MeV. At a given $\mu_q$ the gluon mass is chosen to fit the lattice data points. The impact of the variation of $\alpha_s$ is indicated as the error bands. The vacuum result is also shown for a reference.}
\label{fig:Dg_E_dmu_mg}
\end{figure*}

\begin{figure*}[thbp]
\centering
\includegraphics*[scale=0.52]{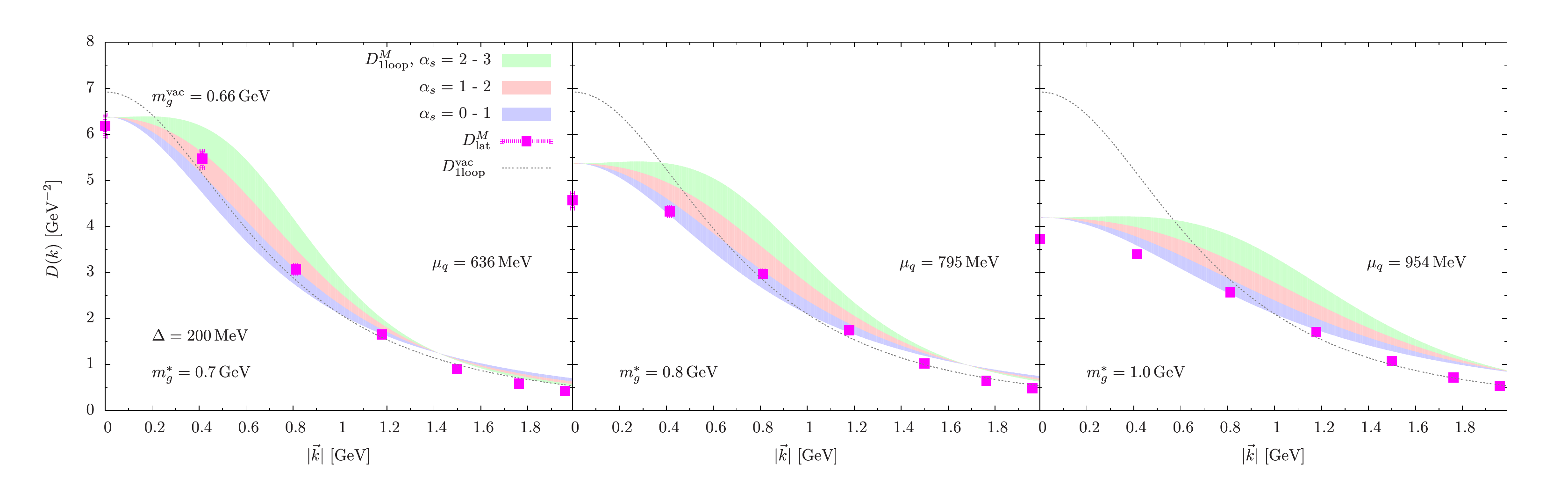}
 \caption{(color online) The magnetic gluon propagators for the setup same as Fig.\ref{fig:Dg_E_dmu_mg}. With larger $\alpha_s$, the polarization effects enhance more the propagators at finite momenta (the {\it para}-magnetic contributions dominate over the {\it dia}-magnetic ones). The difference in $\Delta$ has little impact and is difficult to see in this plot.}
\label{fig:Dg_M_dmu_mg}
\end{figure*}

We have seen in the previous section that the one-loop result is not satisfactory in reproducing the lattice data. A part of the reasons is the lack of the RG improvement which is still within the perturbative framework. Another possible missing piece is non-perturbative modifications of the gluon mass. Such effects should appear when we construct the gap equations for the gluon mass that contain gluon loops \cite{Cornwall:1981zr}.  Solving the gap equations is beyond the scope of this work, but in this section we examine how large the in-medium gluon masses can be.

To examine the possible impacts of non-perturbative modifications of gluon masses, we fit the lattice data in medium by modifying the parameter $m_g$ in the one-loop result from the vacuum value. We write this new parameter $m_g^*$ and optimize its value for each $\mu_q$. We fix the $Z_g^{\rm overall}$ to the vacuum value. The value of $\alpha_s$ is again varied, but this time we classify its error band for a wider range of $\alpha_s$; the domain of $\alpha_s$ is divided into $[0, 1]$, $[1,2]$, and $[2,3]$, and we attach the error band for each. Meanwhile we examine only $\Delta =200$ MeV as it gives better fit than the other choices. As we have found in Figs.\ref{fig:Dg_E_dmu} and \ref{fig:Dg_M_dmu}, the lattice data suggests that the overall structure is similar in the electric and magnetic sectors. For this reason we take the same values for $m_g^*$ in both sectors at given $\mu_q$.

The results for the electric and magnetic sectors are shown in Figs.\ref{fig:Dg_E_dmu_mg} and \ref{fig:Dg_M_dmu_mg}. We found that reasonably good fits are obtained for both electric and magnetic sectors when we choose $m_g^* = ( 0.7, 0.8, 1.0)$ GeV for $\mu_q= 636,\, 745$, and $945$ MeV, respectively. Compared to the vacuum value $m_g^{{\rm vac}} \simeq 0.66$ GeV, at $\mu_q \simeq 1$ GeV its value is enhanced by $\sim 50\%$. One might think this modification is large, but it is much smaller than that expected from the perturbative framework with $m_g=0$. We would say the change is modest.

In principle the evolution of $m_g$ as a function of $\mu_q$ should be determined by solving the gap equations for the gluon mass. While at one-loop the electric and magnetic gluons are protected from the screening effects, we saw that their finite momentum behaviors are not protected from the medium effects.  As the gap equation uses the gluon and quark propagators for all momenta, it is natural to expect that such finite momentum components modify the structure of the gap equation and thereby the resulting non-perturbative gluon mass.

Finally we emphasize again that the growth of $m_g$ makes the dependence on $\alpha_s$ less significant. As discussed in the vacuum case this tendency gives us a hope that proper identification of quasi-particles and the parameters make the residual interactions under control.

\section{summary}
\label{sec:summary}

In this paper we have studied the in-medium gluon propagator in QC$_2$D by employing the CF model
as an effective theory at the energy less than $\sim 1$ GeV. While $\alpha_s$ is known to be large in the infrared, the strong coupling effects may be largely absorbed into the parameters characterizing the quasi-particles in medium, so that the residual interactions may be under control. Our studies in this paper indeed indicates that the gluon mass significantly tempers the $\alpha_s$ corrections, compared to the case of massless gluons. 

The study of quasi-particle picture at $\mu_q=0.5$-$1.0$ GeV is an important step to predict a variety of quantities relevant at the cores of neutron stars where $\mu_q=0.3$-$0.8$ GeV or the corresponding baryon density may reach $\simeq 10n_0$. The pQCD calculations suggested that the matter below $\mu_q\simeq 1$ GeV (or $n_B\lesssim 50\, n_0$) should be regarded as strongly correlated matter. The open question is whether such strongly correlated matter can accommodate quasi-particles or no such simplification occurs. Meanwhile it is not unreasonable to expect the validity of quasi-particle descriptions by referring to the success of constituent quark models in describing the hadron dynamics; inside of hadrons the $\alpha_s$ used for the interaction is $\sim 1$, but the level splitting due to the one-gluon-exchange can capture the overall features of the hadron spectroscopy, provided that the confinement is supplied by collective effects of gluons (rather than their quasi-particle contributions).

The present work must be significantly improved in several respects. First, we need to perform the RG improvement of the one-loop result by using the in-medium gluon propagators inside of the loop graphs; by doing this the electric and magnetic components couple in a nontrivial way and it may explain the similarity of the electric and magnetic sectors in the lattice results. Second, we need to consider the possibility of the non-perturbative modification of the gluon mass by solving the gap equation for the gluon mass; by taking into account the medium modification of finite momentum behaviors, the gap equation itself is modified and so does the solution. Third, we need to estimate the diquark gap, both by performing theoretical calculations and also by extracting the value from the lattice results. Finally, we need to understand better the systematics of the lattice results, the finite volume effects in particular \cite{Fischer:2007pf,Bornyakov:2009ug}; according to the current precision the gluon mass in medium may vary to a factor of two or so. All of these require hard work but seem doable.

Once successful descriptions are established for QC$_2$D, one will be able to utilize the understanding for the quark matter domain to strengthen the constraints for three-color QCD. In particular the equations of state at $n_B \gtrsim 5$-$10\, n_0$ seems calculable but not much work has been done based on the up-to-date frameworks for the non-perturbative physics. Some works toward this direction can be found in Refs.\cite{Klahn:2015mfa,Bai:2017wvk,Song:2019qoh}, but more will be needed to establish our baseline for the phenomenological applications.

\acknowledgments

We thank A. Maas and J. Skullerud for kindly providing us with their lattice data, 
and J. Serreau for very useful comments that helped us to improve the first version of the manuscript considerably.
D.S. is supported by NSFC grant 20201191997.
T.K. is supported by NSFC grant 11875144 and by the KMI for his long-term stay in the Nagoya University.  
D.S. thanks Tetsuo Hatsuda, Mikl\'{o}s Horv\'{a}th, Defu Hou, Dirk H. Rischke, and Igor A. Shovkovy for fruitful discussions.

\end{document}